\documentclass[10pt,journal,compsoc]{IEEEtran}
\makeatletter \def\mdseries@tt{m} \makeatother 

\usepackage{booktabs}
\usepackage{graphicx}
\usepackage[singlespacing]{setspace}
\usepackage{threeparttable}
\usepackage{multirow}
\usepackage{subcaption}
\usepackage[utf8]{inputenc}
\usepackage[english]{babel}
\usepackage{amsthm}
\usepackage{changepage} 
\usepackage{mathtools}
\usepackage{textgreek}
\usepackage[compact]{titlesec}
\usepackage{amsmath}
\usepackage{tikz}
\usetikzlibrary{shapes,positioning,arrows,calc}
\newcommand*\circled[1]{\tikz[baseline=(char.base)]{
            \node[white,shape=circle,draw,inner sep=1pt, fill=blue] (char) {#1};}}
\DeclareFontFamily{OT1}{pzc}{}
\DeclareFontShape{OT1}{pzc}{m}{it}{<-> s * [1.1] pzcmi7t}{}
\DeclareMathAlphabet{\mathpzc}{OT1}{pzc}{m}{it}
\theoremstyle{definition}
\newtheorem{definition}{Definition}
\DeclareMathAlphabet\mathbfcal{OMS}{cmsy}{b}{n}
\usepackage{centernot}
\usepackage{enumitem}
\makeatletter
\def\algbackskip{\hskip-\ALG@thistlm}
\usepackage{xcolor}
\usepackage[ruled, vlined, linesnumbered]{algorithm2e}
\SetAlFnt{\small\sffamily}

\SetCommentSty{mycommfont}
\usepackage{amsmath, listings, amsthm, amssymb, proof, xspace}
\usepackage{hyperref}
\hypersetup{colorlinks,allcolors=black}
\theoremstyle{remark}

\usepackage{forest}
\usetikzlibrary{angles}
\usepackage{array}
\newcolumntype{?}{!{\vrule width 2pt}}
\usepackage{xurl}
\usepackage{authblk}
\usepackage[backgroundcolor=white,bordercolor=blue,linecolor=blue]{todonotes}
\newcommand{\TR}[2]{\textcolor{black}{#2}}
\newcommand{\TRC}[1]{\textcolor{black}{#1}}
\usepackage{tcolorbox}

\author{Ahmadreza Saboor Yaraghi,
        Mojtaba Bagherzadeh,
        Nafiseh Kahani,
        and Lionel Briand, \IEEEmembership{Fellow, IEEE}
\thanks{A. Saboor Yaraghi, M. Bagherzadeh are with School of EECS, University of Ottawa, Ottawa, ON K1N 6N5, Canada.}
\thanks{N. Kahani is with the Department of Systems and Computer Engineering, Carleton University, Ottawa, ON K1S 5B6, Canada.}

\thanks{L. Briand holds shared appointments with the SnT Centre for Security, Reliability and Trust, University of Luxembourg, Luxembourg and the school of EECS, University of Ottawa, Ottawa, ON K1N 6N5, Canada.}
}

\newcommand{\subjectCount}{25 }

\begin{document}

\title{Scalable and Accurate Test Case Prioritization in Continuous Integration Contexts}

\IEEEtitleabstractindextext{
\begin{abstract}

Continuous Integration (CI) requires efficient regression testing to ensure software quality without significantly delaying its CI builds. This warrants the need for techniques to reduce regression testing time, such as Test Case Prioritization (TCP) techniques that prioritize the execution of test cases to detect faults as early as possible. Many recent TCP studies employ various Machine Learning (ML) techniques to deal with the dynamic and complex nature of CI. However, most of them use a limited number of features for training ML models and evaluate the models on subjects for which the application of TCP makes little practical sense, due to their small regression testing time and low number of failed builds. 

In this work, we first define, at a conceptual level, a data model that captures data sources and their relations in a typical CI environment. Second, based on this data model, we define a comprehensive set of features that covers all features previously used by related studies. Third, we develop methods and tools to collect the defined features for 25 open-source software systems with enough failed builds and whose regression testing takes at least five minutes. Fourth, relying on the collected dataset containing a comprehensive feature set, we answer four research questions concerning data collection time, the effectiveness of ML-based TCP, the impact of the features on effectiveness, the decay of ML-based TCP models over time, and the trade-off between data collection time and the effectiveness of ML-based TCP techniques.

\end{abstract}
}

\maketitle

\newcommand{\para}[2]{\textit{\textbf{#1}: {#2} \\}}
\newcommand{\codee}[1]{\small\textit{#1}\normalsize}
\newcommand{\code}[1]{\small\texttt{#1}\normalsize}

\section{Introduction}
\label{sec:introdution}

Application of Continuous Integration (CI) significantly reduces integration problems, speeds up development time, and shortens release time by allowing software developers to integrate their work more frequently with the mainline codebase rather than with deferred integration. Each integration is automatically built and usually validated by regression testing (a CI cycle), upon completion of which the developers are provided feedback. The execution of regression tests may require significant computational resources and take hours or even days to be completed. The prolonged execution of regression tests can delay the CI cycles and prevent timely feedback to the developers.

Test Case Prioritization (TCP) techniques address the long execution time of regression testing by prioritizing (ranking) the execution of test cases such that faults can be detected as early as possible, i.e., the test cases with high fault detection probability and lower execution times are given higher execution priority. These techniques can be classified into heuristic-based and ML-based techniques. Heuristic-based techniques often make use of code coverage analysis and test execution history. Collecting coverage information is in general challenging and, furthermore, precise coverage information requires dynamic analysis, which is difficult to apply in practice, more particularly so in a CI context, mainly due to computational overhead and applicability issues~\cite{elbaum2014techniques,spieker2017reinforcement,lima2020learning,do2020multi,memon2017taming}. Concerning heuristics based on the execution history, relying solely on such history for TCP may not lead to stable results for complex systems in a CI context~\cite{spieker2017reinforcement,ourRL}. In addition to cost and effectiveness issues, heuristics are often defined statically, and there is no standard procedure to tune them based on new changes. Adapting to new changes is critical for TCP in a CI context, due to the frequently-changing codebases.

ML-based TCP techniques train ML models based on various features collected from different sources, such as execution history, to prioritize test cases. In general, ML techniques enable the training of effective models based on imperfect features. They also can adapt to new changes either through incremental learning~\cite{gepperth2016incremental} or retraining. Thus, many researchers have relied on ML techniques to address TCP in the CI context. According to a recent survey~\cite{oursurvey}, various ML-based TCP techniques have been applied in the CI context. However, existing work has not relied on a comprehensive set of features for training ML models, which is crucial to achieve high effectiveness. They also often used inadequate evaluation subjects that have a low number of failed test cases and a very short regression testing time. Evaluating the costs and benefits of a comprehensive set of features cannot be done on such subjects, as applying TCP techniques is practically inefficient in those cases.
Further, none of the existing work reported the cost and time required for collecting features. Thus, despite significant progress, it is still not clear whether or not existing ML-based techniques have reached their full potential for TCP, as they do not take full advantage of all available data sources. Last, several practical questions regarding the application of ML-based techniques remain unanswered, notably what features can be collected and at what cost to support ML-based TCP. These questions include: What is the trade-off between using certain features in terms of data collection time and their impact on the effectiveness of ML models for TCP? How often do the ML models need to be retrained to remain useful? Our goal is to provide concrete recommendations regarding these questions.

To address the issues discussed above, we first define a data model characterizing the operational flow of a typical CI environment (e.g., Travis CI). The data model captures the available data sources and their relations at a high level. We then define a set of features based on the data model and a thorough review of the used features in related work. This results in 150 features across nine groups, which can be collected by analyzing five data sources, including build logs, the source code of test cases and its Version Control System (VCS) history, the code of the system under test, and its VCS history. 

To investigate the benefits and costs of the 150 features, we conducted a large-scale empirical analysis. To do that, we first defined and developed methods to extract the features based on a popular CI tool, Travis CI, and projects written in the Java programming language. We then designed and conducted an extensive experiment, based on the latest 50 builds of 25 subjects with an adequate number of failed test cases and a regression testing time of at least five minutes, to answer the following questions. 
\begin{itemize}
    \item How does data collection time across feature groups compare, and are they significantly different?  \\
    The result shows that data collection time ranges between 0.1 to 11.7 minutes across subjects for each build, the main portion of which is related to features that require static coverage analysis.
    \item How is the effectiveness of ML-based TCP using the defined comprehensive feature set? How does the use of each feature impact effectiveness? \\
    ML model trained using the full comprehensive set of features can reach promising results for TCP for most study subjects. Test execution history features have the greatest impact on TCP effectiveness. 
    \item How often do the ML-based TCP models need to be retrained? What is the best trade-off between retraining frequency and model effectiveness?\\
    The result shows that retraining ML-based TCP models should be performed no less frequently than every 11 builds, and as frequently as possible to achieve the best TCP effectiveness.
    \item What are the trade-offs between the data collection time of features and their impact on the effectiveness of ML-based TCP? \\
    Depending on the acceptable degree of data collection overhead and the need for higher effectiveness, we suggest four alternative choices: using the comprehensive set of features for training the ML model, (1) with or (2) without retaining at each build, (3) using only features based on the execution history of test cases for training, and (4) rely on heuristics based on the failure history of test cases.
    
\end{itemize}

Overall, this work makes the following contributions:
\begin{itemize}
    \item Collection and evaluation of a comprehensive feature set for training ML models for TCP. This set includes all features used by previous studies and addresses all data sources available in the CI context. No previous study is nearly as comprehensive as our work in this respect. 
    \item Answering four important practical questions based on extensive empirical analysis, as discussed above. 
    \item A benchmark of 25 subjects with 21.5k builds and 2.5k failed builds that enables a fair comparison and evaluation of future TCP techniques. We also made our data collection tools available, which can be used to extend and update the subjects \footnote{\underline{\url{https://github.com/Ahmadreza-SY/TCP-CI}}}.
\end{itemize}

The rest of this paper is organized as follows. Section~\ref{sec:background} discusses ML-based TCP in CI context, static coverage analysis, and the classification of source code changes. We review related work in Section~\ref{sec:related-work}. Section~\ref{sec:approch} defines the data and features used in our work and describes how each feature can be practically collected. We then discuss our evaluation approach and results in Section~\ref{sec:validation}, and conclude the paper in Section~\ref{sec:conclusion}.

\section{Background}
\label{sec:background}
\TR{E1 R2.6 R3.1}{In this section, we first describe how ML-based test case prioritization (TCP) fits into a continuous integration (CI) context. We precisely define the ML problem that our study aims to solve. We then explain the methods that are used to extract dependencies between source code entities and test cases and among source code entities. Further, we present our approach for classifying software systems' changes (commits) into \textit{defect-fix} and \textit{non-defect} classes. The dependency and classification of changes are required for computing coverage-based test case features. More specifically, the dependency data is used for the change and impact analysis features, and the classification data is used for test case coverage features. Details regarding the definition of the features are discussed in Section~\ref{sec:approch}.}

\subsection{ML-based Test Case Prioritization in CI Context}
\label{sec:bg-context}
\begin{figure*}[ht!]
    \centering
    \includegraphics[width=0.6\textwidth]{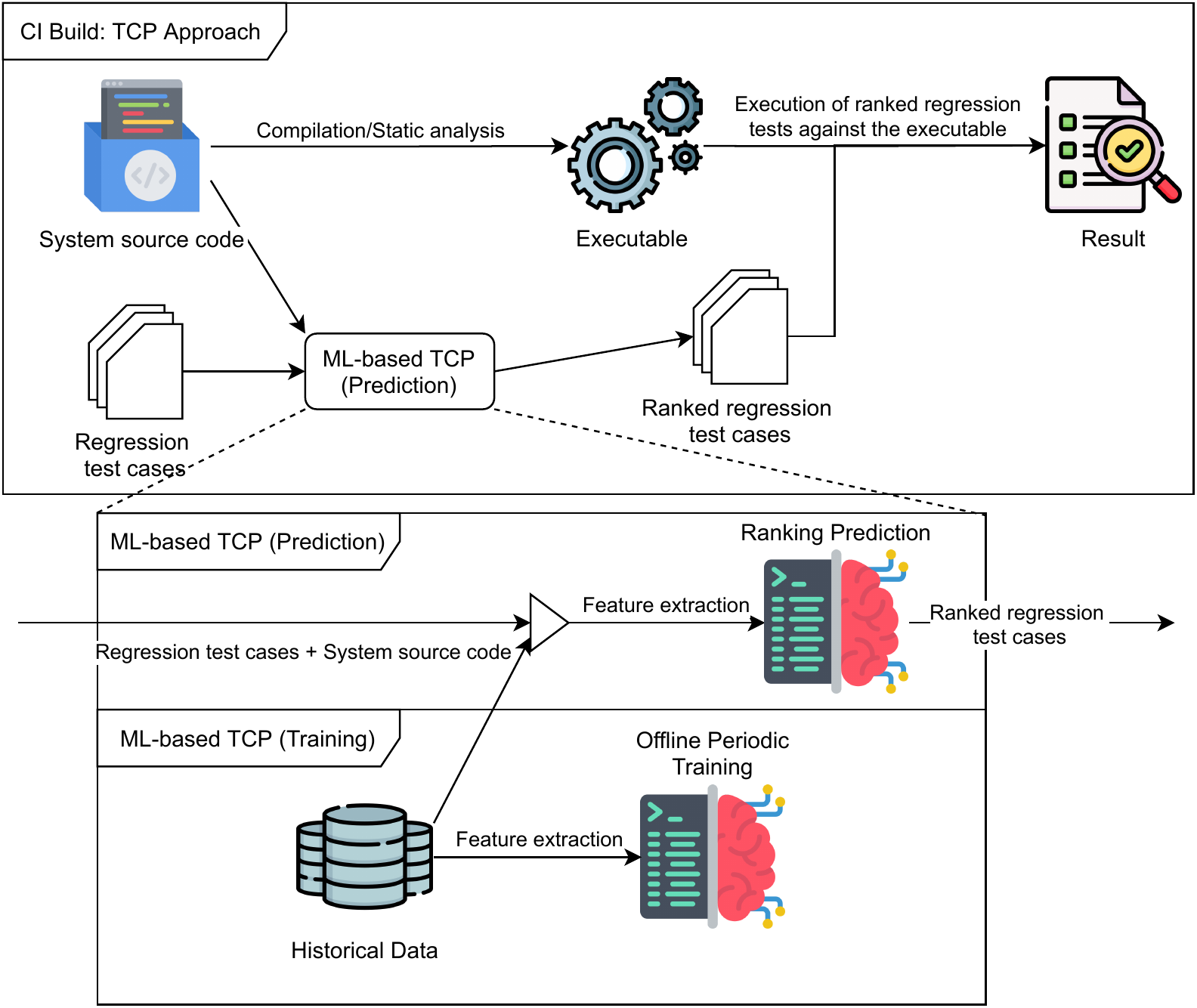}
    \caption{\TRC{ML-based TCP in CI Context}}
    \label{fig:tcp-ci-context}
\end{figure*}

\TR{R2.3 R3.2}{In a typical CI build, the system source code is built, and then its quality is validated by running a set of regression test cases, whose execution can be costly in terms of time and computation resources. Test Case Prioritization (\textit{TCP}) techniques address this challenge by prioritizing (ranking) their execution such that test cases with a higher probability of fault detection and lower execution times are given higher execution priority. Depending on the execution budget, top-ranked test cases can be selected for execution or the regression testing stops once a test case fails. Figure~\ref{fig:tcp-ci-context} depicts a CI build process relying on ML-based TCP. First, similar to a typical CI build, the process begins with building the system. Second, test case features are extracted from several data sources (e.g., system source code) as discussed in Section~\ref{sec:approch}. Third, the extracted features are passed to a pretrained ML ranking model for prediction. This model ranks regression test cases to be executed against the system. The pretrained ML ranking model is trained periodically in an offline environment and therefore does not cause computational overhead to the CI build.}

\TR{R2.9}{The feature extraction step of ML-based TCP, which occurs before ranking regressions test cases for each build, may significantly affect the overall CI runtime. Though the computation of some test case features only entails a simple analysis (e.g., a database query), other features require code analysis and the computation of metrics, thus possibly delaying the execution of a CI build. This motivates us to investigate the time needed for extracting and computing regression test case features for each CI build and to measure the delay it causes. We investigate this issue in the first research question (RQ1) in Section~\ref{sec:validation}.}

\TR{R2.5 R3.2}{Given a test suite $T$ that contains a number of regression test cases, a feature set $F$ that contains features for all test cases, and a ML ranking model $M$, we define an ML-based TCP as a function that takes $M$ and $F$ as input and produces an ordering of $T$ called $T^{\ast}$ that is intended to be as close as possible to the unknown optimal ordering. In this work, similar to existing TCP techniques~\cite{bertolinolearning}, we assume that in the optimal ordering of test suite $T$ ($T^{opt}$), test cases are first sorted by their verdict (i.e., fail/pass state after execution) with failed test cases at the beginning. Second, test cases are sorted by their execution time in ascending order. This ensures the detection of faults as early as possible during regression testing. Also, when the execution budget is limited, the execution of top-ranked test cases (Test Case Selection) may be the only option for an efficient use of the budget.}

\subsection{Dependency and Impact Analysis of Source Code Entities}
\label{sec:dependencyanalysis}

Code coverage can be calculated in different ways depending on the selected analysis techniques. In this work, for practical reasons related in part to scalability as pointed out by the industry partner supporting this research, we use a mix of lightweight static analysis and association rule mining to calculate the dependency graph between source code files of the System Under Test (SUT) and its test cases, assuming that test cases are developed using the same language as the SUT. The dependency graph is a directed graph, each node of which refers to a source file. Each edge shows a dependency relation from the source node to the destination, i.e., the source node depends on the destination. The edges are also weighed with coverage scores calculated based on the association rule mining of co-changes of source files corresponding to the source and destination nodes. In the following, we refer to the source code file corresponding to a node by only referring to the latter.

\TR{R2.14}{A dependency graph is constructed for a version of the SUT and its test cases in a specific build according to the following steps. First, the algorithm that creates the dependency graph accepts a set of test cases and all changes (commits) of the system up to the target version. It then iterates over all source files and creates a node for each one of them. Edges from a node to other nodes are added if the former either calls a function from the latter or imports them. When the graph is completed, then a dependency score is calculated for each edge based on the association rule mining results of co-changes among nodes. The algorithm iterates over all commits of the development history of a given SUT, extracts all pairs of co-changes which are associated with edges of the dependency graph, and finally calculates the support, confidence, and lift for each edge as discussed below.}

Assuming that $CH$ is a \textbf{list} of change sets in the project's commit history: \(CH = (\{f_1, f_2, f_3\}, \{f_1, f_3\}, \{f_2\}, \{f_1, f_2, f_3, f_4\}, ...)\) with $f_i$ representing source files, let us define the following helper functions.
\[ \text{p\_cnt}(f_i, f_j) =  |\{(f_i, f_j) \subseteq E: E \in CH \land 1 \leq i, j \leq n \land i \ne j \}|  \] 
\[ \text{cnt}(f_i) = |\{f_i \in E: E \in CH \land 1 \leq i \leq n \}| \]
where $n$ is the number of source files. Function $\text{p\_cnt}(f_i, f_j)$ computes the number of commits in which $f_i$ and $f_j$ were changed together. Also, function $\text{cnt}(f_i)$ computes the number of commits in which $f_i$ was changed. We define support, confidence, and lift as follows.
\[\text{support}(f_i, f_j)=\frac{\text{p\_cnt}(f_i, f_j)}{|CH|}\]
\[\text{confidence}(f_i, f_j) = \frac{\text{p\_cnt}(f_i, f_j)}{\text{cnt}(f_i)}\]
\[\text{lift}(f_i, f_j) = \frac{\text{p\_cnt}(f_i, f_j)}{\text{cnt}(f_i) * \text{cnt}(f_j)}\]

\begin{definition}{\textit{Coverage}}.\label{def:covarges} Here, we assume that the test cases are developed using the same language as the SUT. Thus, the source files covered by a test case refer to all source files that the test case depends on, i.e., the test case either calls a function in the related source files or imports them. 

The coverage score refers to the confidence of an association rule between a test case and a source file that is computed using association rule mining. To be more precise, the coverage score function (cov\_score) is defined as:
\[\text{cov\_score}(f, t) = \text{confidence}(f, t)\]
where $f$ is a source file and $t$ is a test case. In other words, $\text{cov\_score}(f, t)$ estimates the conditional probability of $t$ being changed given that $f$ is changed.

\end{definition}

\subsection{Classification of Defect-fix Commits}
\label{sec:bg-commitclf}

\begin{definition}{\textit{Previously Detected Faults (PDF)}}.\label{def:pdf} Let us define Previously Detected Faults (PDF) of a source file $f$ as the number of faults detected in $f$ according to its change history:
\[\text{PDF}(f) = |\{c \in C | f \in \text{chn}(c) \land \, \text{chn\_cls}(c)=\text{\textit{defect-fix}} \}|\]
where $C$ is the set of all commits of a project, $\text{chn}(c)$ is a function that returns a set of changed files for commit $c$, and $\text{chn\_cls}(c)$ is a commit classifier that classifies commit $c$ into \textit{defect-fix} or \textit{non-defect} classes.
\end{definition}

The classification of commits is still an active research area, initiated by Mockus et al.~\cite{mockus}, who used three keyword-based rules to automatically classify modification requests' textual descriptions into four maintenance groups. Hindle et al.~\cite{hindle} applied machine learning models using project-dependent features (e.g., authors, modules, and file types) as well as commits' word distributions. Levin et al.~\cite{levin-ds} also used a keyword-based approach as well as source code changes from commits. In a recent effort, Zafar et al.~\cite{zafar-ds} created a commit classifier that reached the high accuracy of 92.2\% for their test dataset. They used BERT~\cite{bert}, a state-of-the-art text classification method. 

BERT~\cite{bert} is a pretrained transformer-based language model which is trained on massive corpora including BooksCorpus~\cite{BooksCorpus} with 800M words and English Wikipedia with 2,500M words. BERT achieved state-of-the-art results on a number of natural language processing tasks. The original $BERT_{BASE}$ model has 110M parameters with 12 encoder layers, and it requires GPU resources for effective training and prediction. For this reason, BERT is computationally expensive and compared to other machine learning models, this is a major hurdle in a CI context, due to the resource and timing constraints of CI builds.

Thus, we aim to use TF-IDF (term frequency-inverse document frequency) and simpler classification techniques (e.g., Random Forests~\cite{random-forests} and SVM~\cite{svm}) to train a commit classifier that requires much less computation time than BERT~\cite{bert}, while retaining high classification accuracy. TF-IDF~\cite{tfidf} measures the importance of a word based on its frequency in a corpus and its presence in individual documents. In the following, we discuss the details of training and evaluating a lightweight commit classifier.   

\textbf{Preparing training datasets.} We collected three publicly available datasets in the literature, including Zafar et al.~\cite{zafar-ds}, Levin et al.~\cite{levin-ds}, and Berger et al.~\cite{berger-ds}, and used their union as the training dataset. Each dataset includes commit messages and their binary labels indicating whether a commit is a defect-fix or not. The training dataset included 3,681 commits, 35.2\% of which were defect-fix commits. We applied a number of preprocessing techniques to clean and normalize the text, which included converting letters to lowercase, replacing URLs with a single token, removing non-alphanumeric characters, removing stop words, and stemming. We then used the TF-IDF technique to convert textual descriptions of commits into vectors. 

\textbf{Training and evaluation of classifiers.}  Based on the vectors of commits' descriptions, we trained three classifiers using  Random Forests~\cite{random-forests}, SVM~\cite{svm}, and XGBoost~\cite{xgboost}. We then evaluated the classifiers using k-fold cross validation with k=5, and XGBoost achieved the best accuracy (83.5\% of correct predictions). Thus, we selected XGBoost to train our commit classifier in this work. 

\textbf{Comparison with the state-of-the-art.} To compare our XGBoost classifier with the one based on BERT (i.e, Zafar et al.~\cite{zafar-ds}), we trained four classifiers based on the four datasets that Zafar et al.~\cite{zafar-ds} experimented with. In Table~\ref{tab:commit-clf}, the BERT column shows BERT's accuracy that was published by Zafar et al.~\cite{zafar-ds}, and the XGBoost column shows the average accuracy based on k-fold cross-validation for our XGBoost classifier. As shown in Table~\ref{tab:commit-clf}, BERT outperforms the XGBoost-based classifier by only a few percentage points. As a result, we can clearly benefit, most particularly in a CI context, from the much lower training cost of XGBoost without significantly sacrificing accuracy.

\begin{table}[t!]
\centering
	\caption{Accuracy of two classification models for defect-fix commit classification across four datasets. BERT refers to the model proposed by Zafar et al.~\cite{zafar-ds}. XGBoost refers to our classification model.}
	\label{tab:commit-clf}
     \begin{tabular}{|c|c|c|c|}
    \hline
    \small \textbf{Dataset} & \textbf{BERT} & \textbf{XGBoost}  \\ \hline
    Zafar et al.~\cite{zafar-ds} & 92.2\% & 89.2\% \\ \hline
    Levin et al.~\cite{levin-ds} & 78.0\% & 75.1\% \\ \hline
    Berger et al.~\cite{berger-ds} & 81.8\% & 81.6\%  \\ \hline
    \small Berger et al.~\cite{berger-ds} (subset~\cite{zafar-ds}) & 91.8\% & 90\%  \\ \hline
    \end{tabular}
\end{table}
\section{Data Model}
\label{sec:approch}

In this section, we first propose a data model based on the regression testing of a CI build. We then describe a feature model aimed at addressing the TCP problem with machine learning and show how our data model relates to the features.

\subsection{High-level class diagram}
\begin{figure}
    \centering
    \includegraphics[width=8cm]{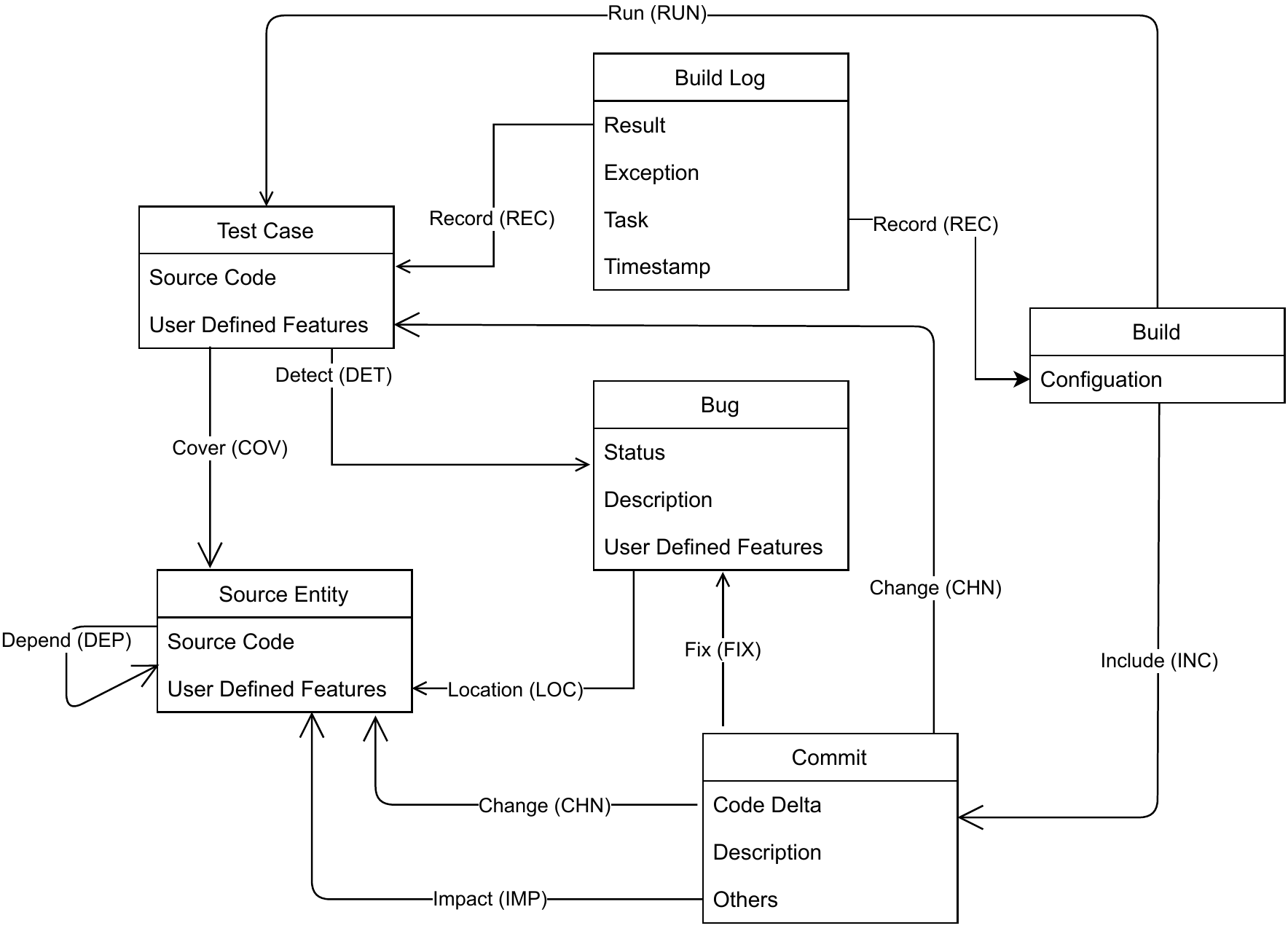}
    \caption{High-level Data Model of a CI Build}
    \label{fig:datamodel}
\end{figure}

Figure~\ref{fig:datamodel} shows a high-level class diagram of entities and properties that are relevant to the regression testing of a CI build. For each CI build, a set of source code files of the SUT are changed (\textit{INC}).  To verify whether or not existing functionalities are impacted by changes, a set of regression test cases are executed (\textit{RUN}), the results of which are captured (\textit{REC}) in the build log. When a test case detects a fault (\textit{DET}), the relevant data is often captured using a fault tracking tool such as Bugzilla. A fault is fixed via changes to the source code (\textit{CHN}) of the SUT and recorded in a code versioning system such as \textit{Git}. A change of source code entity may impact other dependent entities (IMP). Moreover, each test case causes one or more source entities (files or methods) of the SUT to execute (\textit{COV}).

The data represented by our data model is not necessarily available in a structured form and can be available as source code, text files, or even binary files. In the following, we define each of the entities and discuss how their corresponding data can be collected.

\begin{definition}{\textit{Build.}} This data source refers to the CI builds that are defined by scripts and configurations. The scripts are written using a Domain Specific Language (DSL) provided by build tools (e.g., Maven). The scripts specify the details (recipe) of the build, often in the form of rules. A build configuration is a collection of settings (e.g., compiler version) that guides how to run the build scripts. For a CI build to be completed (a build task), the build script needs to be executed based on a specific configuration.  The build scripts and configurations contain metadata required to understand the build logs. Also, build scripts can be analyzed to extract dependencies between source files.

\end{definition}

\begin{definition}{\textit{Build Log.}}
This data source refers to the logs of a CI build. Depending on the underlying build technology, logs may contain different details. Despite such differences, existing build tools typically generate logs that contain the build id, timing information, the result of the build, the output of executed test cases showing whether they passed, and the build configuration (e.g., platform and compiler version). The use of this data requires an analysis of log files that accounts for the specifics of each build tool.
\end{definition}

\begin{definition}{\textit{Test case.}} This data source refers to the source code of test cases and sometimes their descriptions and user-defined properties. 
Such data can be collected via the analysis of the source code of test cases or their descriptions.  Source code analysis typically requires static analysis tools (e.g., Understand~\cite{understand}). Also, natural language processing (NLP) techniques can be applied to extract useful data from source code or test case descriptions.

\end{definition}

\begin{definition}{\textit{Source code.}} This data source refers to the source code of the SUT, which can be accessed and analyzed at three levels of granularity: source file, class, method. Similar to the source of test cases, static analysis is required. 

\end{definition}

\begin{definition}{\textit{Commits.}} This data source refers to all changes to the SUT source code and test cases. A commit captures a change that is applied to a set of source code files. Existing code versioning tools (e.g., Git) provide APIs to access and analyze information about commits.   

\end{definition}

\begin{definition}{\textit{Fault.}} This data source refers to the information about detected faults in the SUT. Existing tools, such as Bugzilla, enable end-to-end tracking of faults that captures when and why a fault is introduced and how it is fixed. In a regression testing context, it is essential to know (2) if a test case reveals a fault during regression testing (regression fault), and (2) how a regression fault is fixed (i.e., commits). While fault tracking tools are widely used, the quality of faults' data is determined by the process followed by development teams to record all relevant details. 

\end{definition}

\subsection{Feature model}
\label{sec:feature-model}
An ML-based TCP model takes feature vectors of test cases as input and ranks these test cases. Thus, for any feature to be used for training, it must be a property of test cases. This often requires aggregating and recasting data collected at a different level of granularity. For example, coverage data shows which source files are covered by which test cases, but such data cannot be directly used to define test case features and must be aggregated and recast as a test case property. 

\begin{figure*}[t!]
    \centering
    \includegraphics[width=0.87\textwidth]{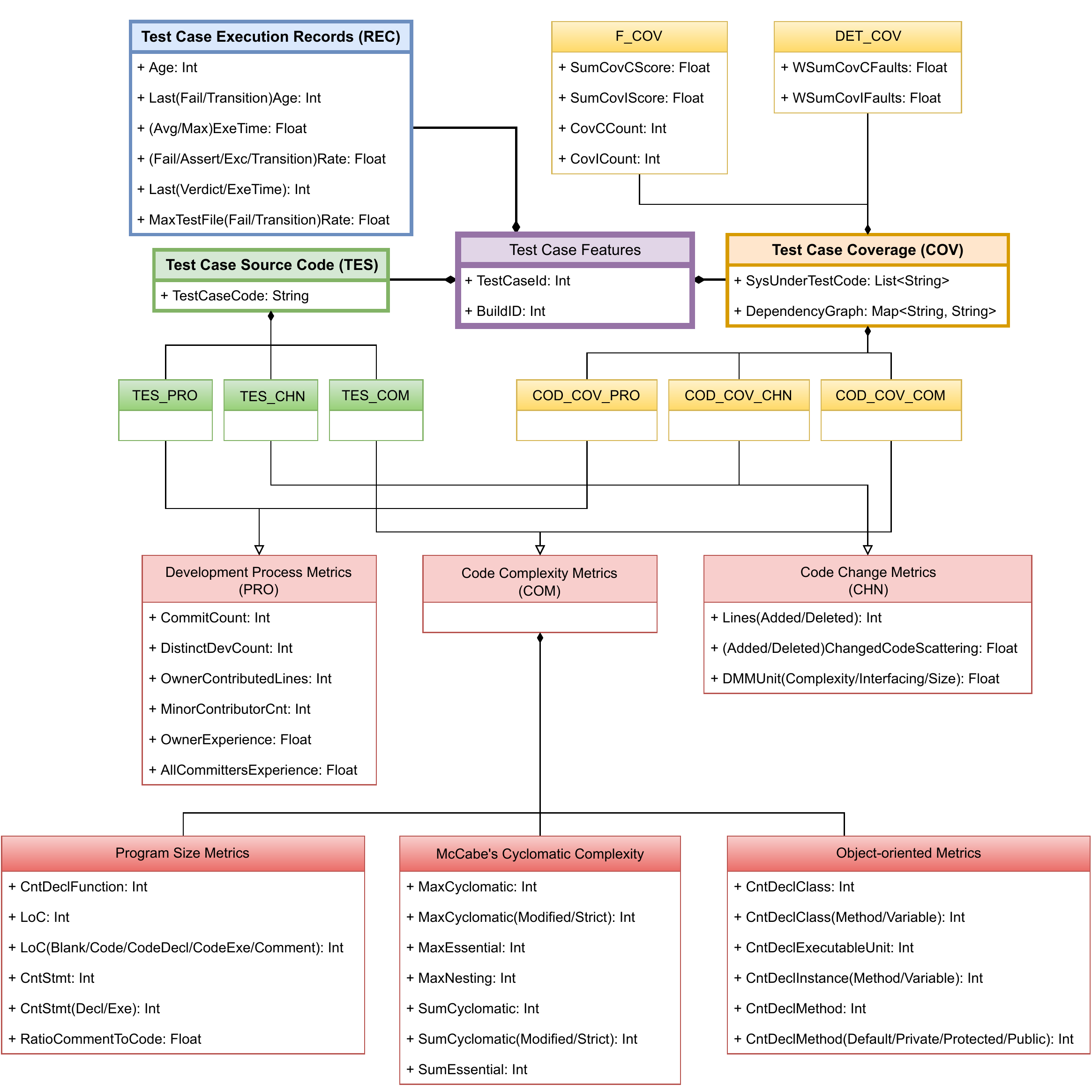}
    \caption{\TRC{A comprehensive class diagram depicting our proposed feature group hierarchy (three high-level groups with nine low-level groups) and the relation between the feature groups and the metrics used for computing them.}}
    \label{fig:features}
\end{figure*}

We define a comprehensive set of test case features that are grouped into three main groups and nine subgroups, as shown in the class diagram in Figure~\ref{fig:features}. In the following, we discuss why each group is considered to be a potentially useful set of features for TCP and how features are calculated based on our data model. \TR{R2.2}{Although most of the features can be calculated at the three different granularity levels (method, class, and file), we only address the file level here because most open-source data sources (e.g., \textit{RTPTorrent}~\cite{RTPTorrent} and \textit{TravisTorrent}~\cite{travisTorrent}) reported test case execution history for test files and classes rather than methods. Coverage features are also collected here at the file level as, for practical reasons invoked by our industry partner, we want to rely on scalable, light-weight static analysis. File-level analysis can however overestimate test case coverage features as each test case file may contain multiple dependent or independent test case methods. A similar analysis and process can be applied for the class and method levels.} Also, we use the naming convention \textit{F\_name} for features, where \textit{name} refers to a meaningful name that is selected based on the source code metrics and the aggregation function (if any) used to calculate the feature. 

Source code metrics that are grouped into three groups (\textit{complexity}, \textit{process}, and \textit{change}) are presented in Figure~\ref{fig:features} and described in the Appendix~\footnote{The Appendix is available as a separate supplementary document.}.
\textit{Complexity} metrics measure the complexity of source code as calculated by static analysis. Also, both \textit{process} and \textit{change} metrics concern how the source code has evolved. However, the former is calculated based on the entire change history of the source code, while the latter is only calculated based on the changes of the latest build. 

Data collection for each metric can be done in two steps: preprocessing and measurement. During preprocessing, all required data are computed and loaded in memory in a certain format (e.g., Abstract Syntax Tree (AST)) that is adequate for the measurement step, during which the metrics are calculated. For instance, to collect complexity metrics for a source file, the file is parsed during preprocessing and represented as an AST, based on which the metrics are calculated. Note that metrics in each group share the same preprocessing step and, as we discuss in Section~\ref{sec:validation}, the cost of the measurement step is significantly less than that of preprocessing. This implies that, in practice, the cost of data collection for a metric in a group is close to that of its entire set of metrics.

To facilitate the definition of features below, let us define functions \textit{chn} and \textit{imp} which take a build $b$ as input and return two disjoint sets of strictly changed and impacted source code files, respectively. The latter correspond to files that may be affected by changes in other files though they were not changed themselves. Results of functions \textit{chn} and \textit{imp} correspond to relations CHN and IMP in the data model (Figure~\ref{fig:datamodel}).

\subsubsection{Test Case Source Code Features (TES)}
These features are calculated based on the complexity, process, and change metrics (Figure~\ref{fig:features}) of the source code files of test cases. The source code features of a test case simply correspond to the metrics of the test case's source file. More details on the definition of metrics can be found in the Appendix. Since this feature group is defined based on source code metrics, we categorize the features in this group into three subgroups, namely TES\_COM, TES\_PRO, and TES\_CHN that correspond to \textit{complexity}, \textit{process}, and \textit{change} metrics, respectively.

The main motivation for using TES features is based on the hypothesis that more complex test cases tend to be associated with longer execution times and a higher probability of detecting faults. Indeed, such test cases are more likely to cover more of the SUT source code and thus have a higher probability of fault detection. Such data are collected through the static analysis of the source code of test cases, which is supported by several tools such as Understand~\cite{understand}, and corresponds to the \textit{Test Case} entity in the data model (Figure~\ref{fig:datamodel}). 

\subsubsection{Test Case Execution Record Features (REC)}
\label{rec-def}

We define test case features which are calculated based on previous execution time records and verdicts (i.e., failed or passed) of the test cases. These features correspond to relation REC between \textit{Test case} and \textit{Build Log} entities in the data model (Figure~\ref{fig:datamodel}).

\begin{itemize}
    \item \textbf{F\_Age:} This feature captures the number of builds from the first execution of the test case (its introduction). Assuming test case $t$ was first executed on build $i$, the age of the test case at build $n$ will be equal to $n - i$. We adopt this feature from previous work~\cite{busjaeger2016learning} which reports that newer test cases fail more often since they exercise new and possibly changed source code.
    
    \item \textbf{F\_LastFailAge:} This feature refers to the number of builds from the last failure of the test case~\cite{newbaseline}. Assuming that the latest failure of test case $t$ occurred on build $i$, \textit{F\_LastFailAge} of $t$ for build $n$ is equal to $n - i - 1$. \textit{F\_LastFailAge} of a test case that has never failed is set to $-1$ rather than $0$, the latter being used for a test case that has failed in the previous build.
    
    \item \textbf{F\_LastTransitionAge}~\cite{newbaseline}: This feature refers to the number of builds since the last change (transition) in the test case's verdict, from failed to passed or vice versa. It is computed in the same way as \textit{F\_LastFailAge} but based on the last transition.
    
    \item \textbf{F\_AvgExeTime:} This feature refers to the average of the previous execution time records of the test case.
    
    \item \textbf{F\_MaxExeTime:} This feature refers to the maximum value of execution time records of the test case.
    
    \item \textbf{F\_FailRate:} This feature is defined as $\frac{f}{n}$, where $f$ is the number of failed executions of the test case and $n$ is the total number of executions of the test case. In this work, we use this rate rather than the failure count of a test case since the latter can be misleading as it is very much dependent on the number of times a test case was executed. 
    
    \item \textbf{F\_AssertRate:} An assertion failure of a test case refers to a failure due to an unexpected output of the SUT. This feature refers to the rate of assertion failures of the test case. 
    
    \item \textbf{F\_ExcRate:} An exception failure is caused by an exception that is not handled correctly in the source code of the test case. This feature refers to the rate of exception failure of a test case. An exception failure may indicate a fault in the source code of $t$ rather than that of the SUT. Thus, we distinguish assertion failures from exception failures.
    
    \item \textbf{F\_TransitionRate}~\cite{newbaseline}: This feature refers to the rate of transitions of the test case verdicts.
    
    \item \textbf{F\_LastVerdict:} This feature captures the verdict of the last execution of a test case.
    
    \item \textbf{F\_LastExeTime:} This feature refers to the last execution time of a test case.
    
    \item \textbf{F\_MaxTestFileFailRate}~\cite{newbaseline}: Assuming that $\text{TFF}(t,f,k)$ refers to the number of builds before build $k$, in which the file $f$ is changed and test case $t$ has failed, and that $\text{TF}(t,k)$ refers to the total number of builds before build $k$ in which test case $t$ has failed, \textit{F\_MaxTestFileFailRate} for test case $t$ in build $k$ with a set of changed files $F$ is calculated as:
    $$
        \frac{\text{Maximum} \; of \; \text{TFF}(t,f,k), f \in F }{\text{TF}(t,k)}
    $$
    If $t$ has never failed, the feature is set to $-1$ rather than $0$, the latter being used when no failure of $t$ co-occurred with changes of the files that are changed in build $k$. The use of this feature is motivated by the fact that, if the previous changes in a file are associated with previous failures of a test case, then the future changes of the file are likely to be associated with failures of the test case.  
    
    \item \textbf{F\_MaxTestFileTransitionRate} \cite{newbaseline}: This feature is exactly computed the same way as \textit{F\_MaxTestFileFailRate} except that it accounts for test case transitions instead of failures.
\end{itemize}

Due to the frequent execution of regression test cases in CI contexts, the volume of execution history is continuously and quickly growing. Thus, \textit{REC} features such as  \textit{F\_AvgExeTime}, \textit{F\_MaxExeTime}, \textit{F\_FailRate}, \textit{F\_AssertRate}, \textit{F\_ExcRate}, and \textit{F\_TransitionRate} are typically calculated based on the latest $n$ test case executions, which are extracted by processing the logs of the latest $n$ builds. Since the main goal of our work is to use a comprehensive set of features, we calculate two values for these six features in our final feature set: \textit{Recent} and \textit{Total}. The \textit{Recent} value is computed based on the latest six builds, similar to previous studies~\cite{bertolinolearning,spieker2017reinforcement}, while the \textit{Total} value is calculated based on all available builds. Previous work~\cite{spieker2017reinforcement} reports that using long text execution history may lead to a reduction in performance. 

The primary motivation for using \textit{REC} features related to fault detection (i.e. \textit{F\_FailRate}, \textit{F\_AssertRate}, \textit{F\_ExcRate}, \textit{F\_TransitionRate}, and \textit{F\_LastVerdict}) is that test cases that detected more faults in the past are more likely to detect faults again in the future, as they tend to exercise complex and frequently changed features. Thus, we conjecture that past failed test cases should be executed again in new builds. Additionally, the hypothesis behind using execution time history is that test cases that take more time to run are more likely to execute more complex and compute-intensive code, as well as larger parts of the code. Therefore, long-running test cases are more likely to detect faults in the SUT. Also, since the execution time of test cases is used as the second criteria for prioritizing test cases (when two test cases have the same verdict, the one with the lower execution time is ranked first), it is an important feature for training ML-based TCP techniques.

\subsubsection{Test Case File Coverage Features (F\_COV)}
These features are calculated based on source files covered (exercised) by the test case and correspond to relation COV in the data model (Figure~\ref{fig:datamodel}). These features capture the ability of test cases to cover source files that are changed (relation CHN) or impacted (relation IMP) in the build.

Let us define function \textit{cov} that takes a source file and a test case as inputs and returns whether the test case covers (exercises) the source file. Also, let us define function \textit{cov\_score} that takes the same inputs and returns the normalized coverage score if the test case covers the source file, and zero otherwise. Since we are eventually going to use coverage scores for weighted summations in builds, we need to use normalized scores. Given a set of coverage scores \(S = \{s_1, ..., s_n\}\) (e.g., coverage scores of all changed files covered by a test case in a build), we normalize them by dividing each score by the sum of all coverage scores. Hence, the normalized coverage score set, $SN$, is defined as follows.
\[SN = \{\frac{s_1}{\sum_{i=1}^{n} s_i}, ..., \frac{s_n}{\sum_{i=1}^{n} s_i}\}\]

The four features in this group refer to the number of covered files and coverage score as defined in the following.

\begin{itemize}
    \item \textbf{F\_SumCovCScore} of a test case $t$ in build $b$ refers to the sum of coverage scores of $t$ w.r.t the changed source files in build $b$: $\sum\limits_{f \in \text{chn}(b)}\text{cov\_score}(f,t)$.
    \item \textbf{F\_SumCovIScore} of a test case $t$ in build $b$ refers to the sum of coverage scores of $t$ w.r.t the impacted source files in build $b$: $\sum\limits_{f \in \text{imp}(b)}\text{cov\_score}(f,t)$.
    \item \textbf{F\_CovCCount} of a test case $t$ in build $b$ refers to the number of covered source files by $t$ that are changed in build $b$: $|\{ f: f \in \text{chn}(b) \land \text{cov}(f,t)\}|$.
   \item \textbf{F\_CovICount} of a test case $t$ in build $b$ refers to the number of covered source files by $t$ that are impacted in build $b$: $|\{ f: f \in \text{imp}(b) \land \text{cov}(f,t)\}|$.
\end{itemize}

Impact analysis is relatively time-consuming and therefore separation of the features based on impacted and changed source files allows us to investigate whether or not including impacted files brings significant benefits. Using such coverage scores as features is motivated by the hypothesis that a test case with higher coverage is more likely to exercise faults, and is, therefore, more likely to detect them.
 
\subsubsection{Features of Covered Source Code by Test cases (COD\_COV)}
This group includes the source code features of the changed and impacted source files (\textit{Source Entity} in the data model in Figure~\ref{fig:datamodel}) covered by the test cases. These features are calculated based on the complexity, process, and change metrics in Figure~\ref{fig:features}. Similar to the TES feature group, we categorize features of this group into three subgroups which are COD\_COV\_COM, COD\_COV\_PRO, and COD\_COV\_CHN, which correspond to \textit{complexity}, \textit{process}, and \textit{change} metrics, respectively.

Since most test cases cover more than one source code entity, we use the normalized weighted sum of metrics based on the coverage score of the test case w.r.t to source files. Also, similar to F\_COV features, we define separate features for changed and impacted source files. For example, the weighted summation of the \textit{LoC} features of covered files of test case $t$ in build $b$ is calculated as follows, assuming that function $\text{loc}(f)$ computes \textit{LoC} for source file $f$,
\[\text{F\_WSumCLoC}= \sum\limits_{f \in \text{chn}(b)}\text{cov\_score}(f,t)*\text{loc}(f)\]
\[\text{F\_WSumILoC} = \sum\limits_{f \in \text{imp}(b)}\text{cov\_score}(f,t)*\text{loc}(f)\]
where \textit{F\_WSumCLoC} and \textit{F\_WSumILoC} refer to the features based on changed and impacted files, respectively. As discussed above, other metrics can replace $\text{loc}(f)$ in the above equations.

The main motivation for using COD\_COV features based on complexity metrics is due to the hypothesis that the cumulative complexity of covered source files by a test case is an indicator of the execution time and fault revealing power of the test case. Also, change metrics, specifically delta maintainability metrics (DMM)~\cite{DeltaMaintainability}, assess the maintainability implications of changes by estimating the risk level of each change. Thus, features defined based on maintainability metrics are good indicators of risky changes, i.e., changes that are likely to be faulty. The same argument applies to features based on process metrics since they indicate the risk entailed by changes by relying on metrics that concern the development process. For example, a change by a new developer has a higher probability of being faulty than that of an experienced developer. Thus, the execution of test cases that cover source files including higher risk changes in the current build has a higher fault detection probability in the context of regression testing.

\subsubsection{Test Case Coverage Fault Detection Features (DET\_COV)} \label{DET} These features are defined based on the Previously Detected Faults (PDF, Definition~\ref{def:pdf}) of source files that is captured by relations DET and LOC in the data model (Figure~\ref{fig:datamodel}). We define the following features in this feature group:
\begin{itemize}
    \item \textbf{F\_WSumCovCFaults} refers to the weighted sum (weighted by coverage scores) of PDFs of the changed source files covered by test case $t$ in build $b$. 
    \[\sum\limits_{f \in \text{chn}(b)}\text{PDF}(f)*\text{cov\_score}(f, t)\]
    \item \textbf{F\_WSumCovIFaults} refers to the weighted sum (weighted by coverage scores) of PDFs of the impacted source files covered by test case $t$ in build $b$.
    \[\sum\limits_{f \in \text{imp}(b)}\text{PDF}(f)*\text{cov\_score}(f, t)\]
\end{itemize}

The main motivation for using DET\_COV features is based on the hypothesis that a source file with a higher PDF is more likely to contain faults in the future. Similarly, a test case that covers files with higher PDF is more likely to detect faults.

\section{Validation}
\label{sec:validation}
This section reports on the experiments we conducted to assess the impact of features on the effectiveness and cost of TCP techniques. We first discuss and motivate four research questions. We then describe the subjects of the study and the evaluation metrics. Finally, we explain our experimental process, present the results, and discuss their practical implications.

\subsection{Research Questions}
\subsubsection{RQ1. Data Collection Time}
\begin{itemize}
    \item \textbf{RQ1.1} How does data collection time across feature groups compare and are they significantly different? 
    \item \textbf{RQ1.2} How does accounting for impacted files affect data collection time for each feature group? 
    \item \textbf{RQ1.3} How do subject size (Source Lines of Code), the number of test cases, and the number of builds affect the data collection time?
\end{itemize}

In a CI context, data collection time is a particularly sensitive issue as time is usually strictly limited for regression testing. A feature can be used to train an ML-based TCP if it can be collected within much less time than that of regression test execution. Thus, we define RQ1 to investigate the data collection time of each feature group according to two modes, based on whether or not impacted files are considered (RQ1.1 and RQ1.2). Also, RQ1.3 investigates how the size of subjects in terms of SLOC (Source Lines of Code), the number of test cases, and the number of builds affect data collection times. In particular,  RQ1.3 investigates how data collection time increases as the size of the subject grows and whether scalability issues can be expected for large systems when collecting features. This RQ focuses on feature groups since all features in each group rely on the same preprocessing, which accounts for most of the data collection time. 

\subsubsection{RQ2. TCP Effectiveness}
\begin{itemize}
    \item \textbf{RQ2.1} How effective is test case prioritization when using the full feature set?
    \item \textbf{RQ2.2} How effective is test case prioritization when impacted files are not considered? 
    \item \textbf{RQ2.3} How does the use of each feature group contribute to the effectiveness of the ML-based TCP?
    \item \textbf{RQ2.4} Which specific feature or subset of features has the highest impact on the effectiveness of ML-based TCP models?
    \item \textbf{RQ2.5} How does the effectiveness of heuristic-based TCP models compare to ML-based models?
\end{itemize}

RQ2.1-2.4 are particularly important in the context of CI as the features used for ML-based TCP models must be minimized, especially features that induce high data collection time, without significantly improving TCP effectiveness. Therefore, the goal is to investigate what TCP effectiveness can be achieved with all features (RQ2.1 and RQ2.2) and what feature groups are most important for TCP effectiveness (RQ2.3). Though the analysis in RQ2.3 is based on feature groups, within feature groups, even those with low impacts, some individual features may have a higher impact on effectiveness than others. This is addressed by RQ2.4. Finally, RQ2.5 compares the results of ML-based TCP with heuristic-based TCP to investigate whether or not using ML brings significant advantages that justify its use.

\subsubsection{RQ3. How often do the ML-based TCP models need to be retrained?}

ML-based TCP models, specifically in the context of CI, need to be retrained regularly based on newly collected data, to better reflect the current status of the system and its history. However, retraining can be expensive and incur delays, and thus we should investigate how the effectiveness of ML-based TCP models decays over time when features are not updated and the models not retrained. This analysis will provide us with insights on how often feature data needs to be collected and used for retraining ML-based TCP.

\subsubsection{RQ4. Trade-off between data collection time and TCP effectiveness for features}

RQ1 and RQ2 separately investigate the data collection time and effectiveness of the TCP model based on individual feature groups. For engineers to make informed decisions in specific contexts, regarding which feature groups should be used for ML-based TCP, a trade-off often needs to be made between data collection times and effectiveness. Thus, RQ4 conducts a comprehensive trade-off analysis with the objective of providing concrete guidelines regarding the use of feature groups, that will hopefully lead to acceptable effectiveness within reasonable time for a specific context.

\subsection{Subjects}
\label{sec:subjects}

\begin{table*}[ht]
\caption{The statistics of the 25 carefully selected subjects. The statistics include source lines of code (SLOC), number of commits, length of the time period between the first and last build, number of all builds and failed builds with the failure rate, average number of test cases per build, and average execution time of all test cases per build.}
\resizebox{\textwidth}{!}{
\begin{tabular}{l|lrrrrrrrrr}
\toprule
$S_{ID}$ &                            Subject &  SLOC & Java SLOC & \# Commits &  Time period (months) & \# Builds &  \# Failed Builds &  Failure Rate (\%) &  Avg. \# TC/Build &  Avg. Test Time (min) \\
\midrule
 $S_{1}$ &                          JMRI/JMRI & 4.56M &     1.05M &      69.3k &                     5 &     1,481 &                65 &                 4 &              4364 &                    25 \\
 $S_{2}$ &                    apache/airavata & 1.46M &      731k &      10.0k &                    15 &       236 &                83 &                35 &                49 &                     6 \\
 $S_{3}$ &              SonarSource/sonarqube &  899k &      398k &      31.8k &                    18 &     4,286 &               230 &                 5 &              1309 &                     6 \\
 $S_{4}$ &                       apache/sling &  695k &      388k &      47.4k &                     7 &     1,403 &               343 &                24 &               189 &                     6 \\
 $S_{5}$ &       camunda/camunda-bpm-platform &  653k &      395k &      20.6k &                    34 &       822 &               125 &                15 &               569 &                    23 \\
 $S_{6}$ &                      facebook/buck &  586k &      384k &      26.3k &                    10 &       846 &               130 &                15 &               663 &                    26 \\
 $S_{7}$ &              apache/shardingsphere &  422k &      165k &      29.6k &                     7 &     1,049 &               123 &                11 &               789 &                    17 \\
 $S_{8}$ &             b2ihealthcare/snow-owl &  373k &      212k &      13.4k &                     2 &       277 &                21 &                 7 &                46 &                    10 \\
 $S_{9}$ &                     Angel-ML/angel &  336k &      204k &       3.0k &                    23 &       308 &               124 &                40 &                33 &                    20 \\
$S_{10}$ &              apache/logging-log4j2 &  313k &      166k &      12.7k &                    13 &       441 &               122 &                27 &               544 &                     8 \\
$S_{11}$ &              eclipse/jetty.project &  282k &      199k &      25.0k &                     2 &       192 &                89 &                46 &               137 &                     6 \\
$S_{12}$ &                  optimatika/ojAlgo &  246k &       84k &       1.6k &                    22 &       254 &                72 &                28 &               136 &                     9 \\
$S_{13}$ &                        yamcs/Yamcs &  229k &      123k &       5.6k &                    24 &       504 &                61 &                12 &               114 &                     6 \\
$S_{14}$ &                     eclipse/steady &  221k &       98k &       2.0k &                    13 &       675 &                51 &                 7 &                81 &                     7 \\
$S_{15}$ &           Graylog2/graylog2-server &  182k &       85k &      22.3k &                    53 &     3,668 &               124 &                 3 &               110 &                    20 \\
$S_{16}$ &                    CompEvol/beast2 &  159k &       83k &       3.0k &                    85 &       415 &               115 &                27 &                65 &                     6 \\
$S_{17}$ &               EMResearch/EvoMaster &  158k &       25k &       4.1k &                     7 &       583 &               109 &                18 &               100 &                    12 \\
$S_{18}$ &                    apache/rocketmq &  135k &      100k &       2.0k &                    16 &       536 &                56 &                10 &               182 &                    17 \\
$S_{19}$ &                   zolyfarkas/spf4j &  125k &       79k &      32.6k &                    37 &       587 &               180 &                30 &               116 &                     7 \\
$S_{20}$ & spring-cloud/spring-cloud-dataflow &  104k &       54k &       3.5k &                     9 &       408 &                19 &                 4 &               115 &                    17 \\
$S_{21}$ &      cantaloupe-project/cantaloupe &   98k &       77k &       4.5k &                    29 &       450 &                65 &                14 &               148 &                    11 \\
$S_{22}$ &                thinkaurelius/titan &   85k &       40k &       5.1k &                    25 &       384 &                41 &                10 &                45 &                    48 \\
$S_{23}$ &                     apache/curator &   84k &       58k &       3.1k &                    21 &       517 &                65 &                12 &               115 &                    67 \\
$S_{24}$ &                 jcabi/jcabi-github &   61k &       32k &       2.8k &                    29 &       788 &                 6 &                 $<0.01$ &               176 &                    14 \\
$S_{25}$ &             eclipse/paho.mqtt.java &   61k &       34k &       1.0k &                    16 &       378 &                77 &                20 &                37 &                    15 \\
\bottomrule
\end{tabular}
}
\label{table:subjects}
\end{table*}

\begin{table}[ht]
\centering
\caption{\TRC{The number of frequent-failing (FF) test cases and subject statistics before (B) and after (A) removing the FF test cases. Subjects with no FF test cases are excluded. The subjects that have less than 50 failed builds after removing FF test cases are shown in bold.}}
\resizebox{\linewidth}{!}{
\begin{tabular}{l|r|rr|rr|rr|rr}
\toprule
         &            & \multicolumn{2}{c|}{\# Failed Builds} & \multicolumn{2}{c|}{Failure Rate (\%)} & \multicolumn{2}{c|}{Avg. \# TC/Build} &  \multicolumn{2}{c}{Avg. Test Time (min)} \\
$S_{ID}$ &  \# FF TCs & B & A                                 & B & A                                  & B & A                                 &  B & A                                      \\
\midrule
 $S_{5}$ &          8 &                        174 &               125 &                         21 &                15 &                        575 &               569 &                             24 &                    23 \\
 $S_{3}$ &          7 &                        299 &               230 &                          6 &                 5 &                       1315 &              1309 &                              7 &                     6 \\
 $S_{7}$ &          6 &                        151 &               123 &                         14 &                11 &                        795 &               789 &                             17 &                    17 \\
 $S_{1}$ &          5 &                         94 &                65 &                          6 &                 4 &                       4368 &              4364 &                             26 &                    25 \\
$S_{17}$ &          2 &                        143 &               109 &                         24 &                18 &                        101 &               100 &                             13 &                    12 \\
$S_{21}$ &          2 &                         70 &                65 &                         15 &                14 &                        149 &               148 &                             12 &                    11 \\
$\mathbf{S_{22}}$ &          2 &                         60 &                \textbf{41} &                         15 &                10 &                         45 &                45 &                             48 &                    48 \\
$\mathbf{S_{24}}$ &          2 &                         76 &                 \textbf{6} &                          9 &                 $<0.01$ &                        174 &               176 &                             13 &                    14 \\
 $S_{4}$ &          1 &                        697 &               343 &                         49 &                24 &                        189 &               189 &                              6 &                     6 \\
 $\mathbf{S_{8}}$ &          1 &                         58 &                \textbf{21} &                         20 &                 7 &                         47 &                46 &                             22 &                    10 \\
$S_{10}$ &          1 &                        231 &               122 &                         52 &                27 &                        545 &               544 &                              8 &                     8 \\
$S_{11}$ &          1 &                        150 &                89 &                         78 &                46 &                        138 &               137 &                              6 &                     6 \\
$S_{13}$ &          1 &                         72 &                61 &                         14 &                12 &                        115 &               114 &                              7 &                     6 \\
$S_{19}$ &          1 &                        277 &               180 &                         47 &                30 &                        117 &               116 &                              7 &                     7 \\
$\mathbf{S_{20}}$ &          1 &                         88 &                \textbf{19} &                         21 &                 4 &                        115 &               115 &                             17 &                    17 \\
$S_{23}$ &          1 &                        103 &                65 &                         19 &                12 &                        115 &               115 &                             68 &                    67 \\
\bottomrule
\end{tabular}
}
\label{table:changed-subjects}
\end{table}

We ran our experiments on \subjectCount subjects, which were selected in a 6-step process, as discussed in the following.

\begin{enumerate}
    \item We started with the latest available database of open-source projects from \textit{GHTorrent}~\cite{ghtorrent} (dump 2021-03-06 with more than 100 million projects). We then selected active (i.e., not forked and deleted) and popular (with at least 50 stars) Java projects from the database that resulted in 22,551 projects. \textit{GHTorrent} provides regularly updated databases of GitHub open-source repositories along with tooling to search in the databases of active (i.e., not forked or deleted) projects in GitHub. 
    \item We selected projects with at least 100 CI runs hosted on Travis CI that resulted in 3,323 projects. We then used \textit{TravisTorrent}~\cite{travisTorrent} to fetch CI build logs and commits of the selected open-source repositories. \textit{TravisTorrent} provides scripts for fetching data from Travis CI and GitHub (for build commits).
    \item We selected projects that use \textit{Maven} as their test execution tool since it provides build logs that contain the required information regarding test case executions (i.e., verdict, duration, and class name). This further reduced the number of projects to 1,419. 
    \item From the resulting projects in step 3, we selected the union of the top 300 projects with the highest SLOC (Source Lines of Code) and the top 300 projects with the highest number of builds. This resulted in 434 projects.
    \item We used \textit{TravisTorrent}~\cite{travisTorrent} to extract the build logs and build commits of the projects resulting from step 4. We analyzed the build logs to calculate the regression testing duration and failed builds of projects. We then selected projects with at least 5 minutes of average regression testing time and 50 failed builds, which resulted in 18 projects. There is no or little practical value in performing TCP when the regression testing time is less than 5 minutes, as most TCP techniques often require more than a few minutes for the data collection and prioritization of test cases. Also, we require at least 50 failed builds to create sufficiently balanced datasets.
    \item  We selected another 7 projects with at least 5 minutes of average regression testing time and 50 failed builds from the 20 open-source projects provided by RTPTorrent's~\cite{RTPTorrent}, a public dataset for the evaluation of TCP techniques. Thus, overall we selected \subjectCount projects as the subjects of our studies that are shown in Table~\ref{table:subjects}.
\end{enumerate}

\TR{R2.8}{We investigated the failure frequency of test cases among failed builds. For most of the subjects, some of the test cases failed frequently across failed builds. We then investigated the reasons for the existence of such test cases by going through the build logs of a sample of the subjects and reading the error messages caused by test case failures. In the investigation sample, which included $S_{24}$, $S_{20}$, $S_{8}$, and $S_{7}$, we found test cases that failed in more than 65\% of the failed builds due to the same reason. Such reasons included external exceptions which were not related to the SUT, such as errors in HTTP requests to external APIs due to authentication issues, invalid arguments, or unexpected responses. Common failure causes also included Java runtime errors, such as \textit{ClassNotFoundException} or \textit{FileNotFoundException}, which were due to missing classes or missing files in the project. Executing such frequently-failing (FF) test cases, which are also referred to as \textit{known breakages}, has no practical value in regression testing as they tend to fail most of the time for the same reason, independently of changes. Since our focus is regression testing, we removed these test cases by performing outlier tests, with respect to failure frequency, using the three-sigma rule of thumb\cite{wiki:three-sigma}. As suggested by our analysis, outlier test cases are highly likely to correspond to non-regression failures and therefore tend to blur our empirical results. Note that though such outlier tests may not remove all of FF test cases, we are confident that most of them were identified and excluded from our datasets. Hence, the subject statistics, experiments, and results presented throughout the rest of the paper are based on data in which the \textit{FF} test cases are removed.}

The final \subjectCount subjects were selected by analyzing more than 20,000 popular open-source Java projects. The selection process assures that all subjects have an acceptable number of failed test cases and regression testing time, both of which are critical for the application and evaluation of TCP techniques. Table~\ref{table:subjects} shows the characteristics of the subjects in terms of line of codes, the number of (failed) builds, failure rate, commits, and the average regression testing time per build. The first column (\textit{$S_{ID}$}) of the table shows the identifier of subjects that will be used to refer to them in the rest of this section. 

Column \textit{SLOC} of Table~\ref{table:subjects} shows the total number of code and comment lines of the subjects that were counted based on the latest build. \textit{SLOC} ranges from 61k to 4.56M, with a median of 229k. Compared with the subjects that are used in most recent studies, our work relies on a high number of subjects (25) whose median SLOC is 229k compared to 37.4k ~\cite{bertolinolearning} and 132k ~\cite{newbaseline}, respectively.
Also, the average number of test cases per build across our subjects ranges from 33 to 4368, with a median of 117, which is similar to 18 previous studies reported by~\cite{oursurvey}. Thus, compared to the previous studies, we use a higher number of subjects whose size in terms of SLOC can be considered to be reasonably larger. 
A build can fail for several reasons, including compilation errors or a test case failure. However, in a TCP context, we are only interested in the latter, and failed builds are, in our subjects, builds with at least one failed test case. Column \textit{\# Failed Builds} of Table~\ref{table:subjects} shows the number of failed builds of each subject. This number ranges from 6 to 343, with a median of 83, representing a diverse set of subjects allowing us to conduct a large number of experiments to account for randomness and draw statistically valid conclusions. Pan et al.~\cite{oursurvey} reports that previous studies evaluate their work mainly based on subjects with a low number of failed builds that ranges between 1 and 70 with a median of 9, which results in unbalanced training datasets. Also, it is not meaningful to evaluate TCP techniques based on a subject with a very low number of failed builds as the main goal of TCP is the early detection of faults and most of the evaluation metrics are based on counting detected faults. 

The last column of Table~\ref{table:subjects} shows the average of regression testing time per build across all subjects that ranges from 6 to 67 minutes with a median of 12. Compared to recent studies where 11 out of 23 subjects~\cite{newbaseline} have regression testing times below 3 minutes, or all subjects have regression testing times below 30 seconds~\cite{bertolinolearning}, our subjects are significantly better. Recall that it is not meaningful to apply and evaluate TCP techniques in the context of projects whose regression testing times are small. 

\TR{R2.8}{As discussed earlier, we removed the \textit{FF} test cases from the subjects using the outlier test. Table~\ref{table:changed-subjects} compares the statistics of subjects, which have at least one \textit{FF} test case, before and after removing their \textit{FF} test cases. Column \textit{\# FF TCs} corresponds to the number of \textit{FF} test cases, and the rest of the columns show the statistics before (B) and after (A) removing all FF test cases. As visible, the number of failed builds for some subjects significantly drops when the FF test cases are removed. More specifically, for four of the subjects ($S_{8,20,22,24}$), the number of failed builds decreases to under 50 (our selection criterion) after removing \textit{FF} test cases. This suggests that \textit{FF} test cases were the main cause of build failures across these subjects. However, the effect of removing \textit{FF} test cases on the average number of test cases per build and the average regression testing time per build is negligible.}

\subsection{Evaluation Metrics}
\label{sec:metrics}

In this work, we use Cost-cognizant Average Percentage of Faults Detected ($\text{APFD}_C$)~\cite{elbaum2001incorporating} as the evaluation metric to measure the effectiveness of prioritization techniques. $\text{APFD}_C$ is a cost-aware variant of the well-known and widely used APFD~\cite{rothermel1999test} metric. APFD only measures the extent to which a certain ranking reveals faults early and does not take the execution time of test cases into account, which is important, especially in a CI context. 
Similar to prior work~\cite{newbaseline, project-list-2, chen2018optimizing}, since fault severity information is not available, we assume all faults have the same severity. Also, since our collected data does not include the mapping of faults and test cases, similar to prior work~\cite{newbaseline, project-list-2, chen2018optimizing}, we assume that each test case failure refers to a distinct fault in the SUT. In practice, obviously, this widely used assumption is not correct. However, we can expect the number of faults detected to be roughly proportional to the number of failures.

$\text{APFD}_C$ of a test case ordering $T^{\ast}$ is calculated as:

$$
\text{APFD}_C = \frac{ \sum_{i=1}^{m} {(\sum_{j=TF_i}^{n}{t_j} - \frac{1}{2}t_{TF_i})} }{ \sum_{j=1}^{n}{t_j} \times m}
$$
where $m$ refers to the total number of faults, $n$ refers to the total number of test cases in $T^{\ast}$, and $TF_i$ refers to the position (starting from 1) of the first failed test case in $T^{\ast}$ that detects the $i$th fault, and $t_j$ refers to the execution time of the $j$th test.

Both APFD and $\text{APFD}_C$ can only be computed for builds that contain failures. Thus, here we only report $\text{APFD}_C$ based on failed builds.

\subsection{Experiment Design, Results, and Discussion}
\label{sec:experiments}

\subsubsection{Data collection Time (RQ1)}

\begin{table*}[ht]
\caption{Average preprocessing (P), measurement (M), and total (T) data collection time (in seconds) for all feature groups across subjects. For each column, the maximum value is shown in bold.}
\resizebox{\textwidth}{!}{%
\begin{tabular}{l|rrr|rrr|rrr|rrr|rrr|rrr|rrr|rrr|rrr}
\toprule
        & \multicolumn{3}{c|}{COD\_COV\_COM} & \multicolumn{3}{c|}{COD\_COV\_PRO} & \multicolumn{3}{c|}{DET\_COV} & \multicolumn{3}{c|}{COD\_COV\_CHN} & \multicolumn{3}{c|}{F\_COV} & \multicolumn{3}{c|}{TES\_COM} & \multicolumn{3}{c|}{TES\_PRO} & \multicolumn{3}{c|}{REC} & \multicolumn{3}{c}{TES\_CHN} \\
$S_{ID}$ & P             & M             & T             & P             & M             & T             & P         & M         & T       & P          & M        & T         & P       & M   & T       & P        & M       & T       & P        & M       & T       & P      & M      & T     & P      & M        & T        \\
\midrule
 $S_{1}$ &         \textbf{508.9} &           \textbf{0.4} &         \textbf{509.3} &         \textbf{455.8} &           \textbf{5.5} &         \textbf{461.3} &     \textbf{452.8} &       \textbf{4.1} &     \textbf{456.9} &         \textbf{451.4} &           6.1 &         \textbf{457.5} & \textbf{451.4} &   0.0 & \textbf{451.4} &      \textbf{57.5} &       \textbf{0.1} &      \textbf{57.6} &       4.4 &      \textbf{28.9} &      \textbf{33.3} &   \textbf{0.4} & \textbf{139.1} & \textbf{139.4} &       0.0 &       \textbf{5.6} &       \textbf{5.6} \\
 $S_{2}$ &          61.6 &           0.0 &          61.6 &          23.9 &           0.2 &          24.2 &      23.6 &       0.1 &      23.7 &          23.3 &           2.2 &          25.5 &  23.3 &   0.0 &  23.3 &      38.3 &       0.0 &      38.3 &       0.6 &       0.4 &       1.0 &   0.0 &   0.5 &   0.5 &       0.0 &       0.4 &       0.4 \\
 $S_{3}$ &         262.7 &           0.1 &         262.8 &         244.2 &           0.8 &         245.0 &     244.1 &       0.5 &     244.6 &         243.8 &           4.5 &         248.4 & 243.8 &   0.0 & 243.8 &      18.9 &       0.0 &      18.9 &       0.4 &       5.1 &       5.5 &   0.1 &  21.9 &  22.0 &       0.0 &       4.2 &       4.2 \\
 $S_{4}$ &          98.6 &           0.0 &          98.6 &          74.3 &           0.2 &          74.5 &      74.1 &       0.1 &      74.3 &          74.1 &           1.1 &          75.2 &  74.1 &   0.0 &  74.1 &      24.6 &       0.0 &      24.6 &       0.2 &       0.9 &       1.2 &   0.0 &   1.3 &   1.4 &       0.0 &       1.1 &       1.1 \\
 $S_{5}$ &         130.0 &           0.1 &         130.1 &         107.9 &           0.7 &         108.6 &     107.7 &       0.4 &     108.1 &         106.8 &           1.3 &         108.1 & 106.8 &   0.0 & 106.8 &      23.2 &       0.0 &      23.2 &       1.1 &       3.6 &       4.7 &   0.1 &   6.6 &   6.7 &       0.0 &       1.2 &       1.2 \\
 $S_{6}$ &          69.7 &           0.1 &          69.9 &          49.4 &           0.7 &          50.1 &      49.2 &       0.3 &      49.5 &          48.9 &           1.6 &          50.5 &  48.9 &   0.0 &  48.9 &      20.9 &       0.0 &      20.9 &       0.5 &       3.3 &       3.8 &   0.1 &   5.6 &   5.7 &       0.0 &       1.5 &       1.5 \\
 $S_{7}$ &         110.3 &           0.1 &         110.3 &          93.8 &           0.6 &          94.3 &      93.2 &       0.3 &      93.5 &          92.5 &           1.6 &          94.1 &  92.5 &   0.0 &  92.5 &      17.7 &       0.0 &      17.8 &       1.2 &       3.3 &       4.5 &   0.1 &   6.2 &   6.3 &       0.0 &       1.5 &       1.5 \\
 $S_{8}$ &          47.5 &           0.0 &          47.5 &          33.5 &           0.3 &          33.8 &      32.8 &       0.2 &      33.0 &          31.2 &           0.7 &          31.9 &  31.2 &   0.0 &  31.2 &      16.2 &       0.0 &      16.2 &       2.3 &       0.5 &       2.8 &   0.0 &   0.6 &   0.6 &       0.0 &       0.6 &       0.6 \\
 $S_{9}$ &          29.4 &           0.0 &          29.4 &          21.0 &           0.1 &          21.1 &      20.8 &       0.1 &      20.9 &          19.6 &           0.3 &          20.0 &  19.6 &   0.0 &  19.6 &       9.8 &       0.0 &       9.8 &       1.4 &       0.2 &       1.6 &   0.0 &   0.3 &   0.3 &       0.0 &       0.3 &       0.3 \\
$S_{10}$ &          41.0 &           0.0 &          41.1 &          30.2 &           0.2 &          30.4 &      30.1 &       0.1 &      30.1 &          29.8 &           0.5 &          30.2 &  29.8 &   0.0 &  29.8 &      11.2 &       0.0 &      11.3 &       0.5 &       2.5 &       3.0 &   0.1 &   3.7 &   3.8 &       0.0 &       0.4 &       0.4 \\
$S_{11}$ &          65.3 &           0.0 &          65.3 &          52.2 &           1.1 &          53.2 &      52.0 &       0.8 &      52.9 &          51.6 &           0.6 &          52.3 &  51.6 &   0.0 &  51.6 &      13.7 &       0.0 &      13.7 &       0.5 &       1.7 &       2.2 &   0.1 &   1.4 &   1.4 &       0.0 &       0.6 &       0.6 \\
$S_{12}$ &          22.4 &           0.0 &          22.4 &          16.2 &           0.3 &          16.6 &      16.0 &       0.1 &      16.1 &          14.3 &           0.9 &          15.1 &  14.3 &   0.0 &  14.3 &       8.1 &       0.0 &       8.1 &       2.0 &       0.8 &       2.8 &   0.0 &   1.1 &   1.1 &       0.0 &       0.5 &       0.5 \\
$S_{13}$ &          59.0 &           0.0 &          59.0 &          16.8 &           0.3 &          17.1 &      16.4 &       0.1 &      16.5 &          14.9 &          \textbf{10.4} &          25.3 &  14.9 &   0.0 &  14.9 &      44.1 &       0.0 &      44.1 &       1.9 &       0.7 &       2.6 &   0.0 &   1.0 &   1.1 &       0.0 &       1.0 &       1.0 \\
$S_{14}$ &          70.5 &           0.0 &          70.5 &          59.4 &           0.2 &          59.6 &      59.0 &       0.0 &      59.1 &          50.3 &           0.6 &          50.9 &  50.3 &   0.0 &  50.3 &      20.2 &       0.0 &      20.2 &       9.1 &       0.5 &       9.5 &   0.1 &   0.6 &   0.6 &       0.0 &       0.4 &       0.4 \\
$S_{15}$ &           7.1 &           0.0 &           7.1 &           4.5 &           0.2 &           4.7 &       4.4 &       0.2 &       4.6 &           4.2 &           0.8 &           5.0 &   4.2 &   0.0 &   4.2 &       2.9 &       0.0 &       2.9 &       0.2 &       0.4 &       0.7 &   0.0 &   0.4 &   0.5 &       0.0 &       0.8 &       0.8 \\
$S_{16}$ &          12.9 &           0.0 &          12.9 &           6.3 &           0.1 &           6.4 &       6.2 &       0.0 &       6.3 &           6.2 &           0.2 &           6.4 &   6.2 &   0.0 &   6.2 &       6.7 &       0.0 &       6.7 &       0.2 &       0.4 &       0.6 &   0.0 &   0.6 &   0.6 &       0.0 &       0.1 &       0.1 \\
$S_{17}$ &           6.0 &           0.0 &           6.0 &           2.8 &           0.1 &           2.9 &       2.7 &       0.0 &       2.8 &           2.5 &           0.2 &           2.7 &   2.5 &   0.0 &   2.5 &       3.5 &       0.0 &       3.5 &       0.3 &       0.5 &       0.8 &   0.0 &   1.1 &   1.1 &       0.0 &       0.2 &       0.2 \\
$S_{18}$ &          29.8 &           0.0 &          29.8 &          18.4 &           0.3 &          18.7 &      18.3 &       0.0 &      18.3 &          17.7 &           0.4 &          18.1 &  17.7 &   0.0 &  17.7 &      12.1 &       0.0 &      12.1 &       0.7 &       0.9 &       1.6 &   0.1 &   1.3 &   1.4 &       0.0 &       0.3 &       0.3 \\
$S_{19}$ &          21.1 &           0.0 &          21.1 &          13.6 &           0.7 &          14.3 &      12.6 &       0.5 &      13.1 &          12.5 &           1.6 &          14.0 &  12.5 &   0.0 &  12.5 &       8.6 &       0.0 &       8.6 &       1.2 &       1.1 &       2.3 &   0.0 &   1.6 &   1.6 &       0.0 &       1.5 &       1.5 \\
$S_{20}$ &          23.9 &           0.0 &          23.9 &          29.0 &           0.2 &          29.2 &      27.3 &       0.1 &      27.3 &          16.9 &           3.6 &          20.5 &  16.9 &   0.0 &  16.9 &       7.0 &       0.0 &       7.0 &      \textbf{12.1} &       0.6 &      12.7 &   0.0 &   1.0 &   1.0 &       0.0 &       3.5 &       3.5 \\
$S_{21}$ &          10.1 &           0.0 &          10.1 &           4.5 &           0.2 &           4.7 &       4.4 &       0.0 &       4.4 &           4.2 &           0.3 &           4.6 &   4.2 &   0.0 &   4.3 &       5.9 &       0.0 &       5.9 &       0.2 &       0.8 &       1.0 &   0.0 &   1.3 &   1.3 &       0.0 &       0.3 &       0.3 \\
$S_{22}$ &          10.6 &           0.0 &          10.7 &           8.7 &           0.3 &           9.0 &       8.4 &       0.2 &       8.6 &           6.0 &           0.5 &           6.6 &   6.0 &   0.0 &   6.0 &       4.6 &       0.0 &       4.6 &       2.7 &       0.5 &       3.2 &   0.0 &   0.6 &   0.6 &       0.0 &       0.5 &       0.5 \\
$S_{23}$ &          10.3 &           0.0 &          10.3 &           5.4 &           0.1 &           5.5 &       5.3 &       0.0 &       5.4 &           5.3 &           0.1 &           5.4 &   5.3 &   0.0 &   5.3 &       5.0 &       0.0 &       5.1 &       0.1 &       0.5 &       0.6 &   0.0 &   0.7 &   0.7 &       0.0 &       0.1 &       0.1 \\
$S_{24}$ &           7.3 &           0.0 &           7.3 &           5.2 &           0.1 &           5.3 &       5.1 &       0.1 &       5.2 &           5.0 &           0.2 &           5.1 &   5.0 &   0.0 &   5.0 &       2.3 &       0.0 &       2.3 &       0.2 &       0.7 &       1.0 &   0.0 &   1.3 &   1.3 &       0.0 &       0.1 &       0.1 \\
$S_{25}$ &           6.7 &           0.0 &           6.7 &           3.2 &           0.1 &           3.3 &       3.1 &       0.0 &       3.1 &           2.8 &           0.1 &           2.9 &   2.8 &   0.0 &   2.8 &       3.8 &       0.0 &       3.8 &       0.3 &       0.2 &       0.6 &   0.0 &   0.3 &   0.4 &       0.0 &       0.1 &       0.1 \\ \hline
     Avg &          83.9 &           0.0 &          84.0 &          68.5 &           0.5 &          69.0 &      68.1 &       0.3 &      68.4 &          67.3 &           1.6 &          68.9 &  67.3 &   0.0 &  67.3 &      16.6 &       0.0 &      16.6 &       1.2 &       2.3 &       3.5 &   0.1 &   7.2 &   7.3 &       0.0 &       1.3 &       1.3 \\
\bottomrule
\end{tabular}
}
\label{table:avg-time-subjects}
\end{table*}

\begin{table}[ht]
\caption{Average execution time of all test cases per build (\textit{Avg. Testing Time}) compared to the average data collection time for all feature groups per build (\textit{Avg. Data Collection Time}) across all subjects. The times are reported in minutes, and the maximum value of each column is shown in bold.}
\label{tab:testing-vs-collection-time}
\resizebox{\linewidth}{!}{%
\begin{tabular}{l|rrr}
\toprule
$S_{ID}$ & Avg. Testing Time & Avg. Data Collection Time &  Collection/Testing (\%) \\
\midrule
 $S_{1}$ &              26.6 &                      \textbf{11.7} &                       44 \\
 $S_{2}$ &               6.0 &                       1.1 &                       18 \\
 $S_{3}$ &               7.0 &                       5.0 &                       \textbf{71} \\
 $S_{4}$ &               6.9 &                       1.7 &                       25 \\
 $S_{5}$ &              24.7 &                       2.4 &                       10 \\
 $S_{6}$ &              26.3 &                       1.4 &                        5 \\
 $S_{7}$ &              17.6 &                       2.1 &                       12 \\
 $S_{8}$ &              22.4 &                       0.9 &                        4 \\
 $S_{9}$ &              20.7 &                       0.5 &                        3 \\
$S_{10}$ &               8.1 &                       0.8 &                       10 \\
$S_{11}$ &               6.6 &                       1.2 &                       18 \\
$S_{12}$ &               9.5 &                       0.5 &                        5 \\
$S_{13}$ &               7.5 &                       1.2 &                       16 \\
$S_{14}$ &               7.7 &                       1.4 &                       18 \\
$S_{15}$ &              20.7 &                       0.2 &                        1 \\
$S_{16}$ &               6.3 &                       0.2 &                        4 \\
$S_{17}$ &              13.3 &                       0.1 &                        1 \\
$S_{18}$ &              17.6 &                       0.6 &                        3 \\
$S_{19}$ &               7.1 &                       0.5 &                        7 \\
$S_{20}$ &              17.7 &                       0.7 &                        4 \\
$S_{21}$ &              12.3 &                       0.2 &                        2 \\
$S_{22}$ &              48.8 &                       0.3 &                        1 \\
$S_{23}$ &              \textbf{68.1} &                       0.2 &                        $<1$ \\
$S_{24}$ &              13.7 &                       0.2 &                        1 \\
$S_{25}$ &              15.5 &                       0.1 &                        1 \\
\bottomrule
\end{tabular}
}
\end{table}

\textbf{Overview} Table~\ref{table:avg-time-subjects} reports the preprocessing (sub-column $P$) and measurement (sub-column $M$) times of feature groups across subjects. For computing each test case feature in a CI build, a number of preprocessing steps are required including static source code and dependency analysis, source code change history collection, and text classification. The preprocessed data is used for one or more feature groups. The measurement time refers to the computation of feature values using the preprocessed data. Most of the measurement times, for all feature groups, are less than a second and therefore negligible. In contrast, preprocessing is expensive, taking for many feature groups almost all the data collection time. Such groups require static source code and dependency analysis as well as textual classification. Therefore, in most production environments where systems tend to be as large or larger than the largest system we consider here ($S_1$), it might be more practical to do the preprocessing periodically rather than on each build, as we will discuss in RQ3. For this RQ, however, we performed the preprocessing for all builds, to account for the worst-case situation in terms of data collection time.

Also, Table~\ref{table:avg-time-subjects} shows the average data collection times of feature groups for all builds, across subjects. The first result of practical importance is that the data collection times of certain feature groups are significantly higher than others: COD\_COV\_COM, COD\_COV\_PRO, DET\_COV, COD\_COV\_CHN, and F\_COV with total data collection time averages of 84, 69, 68.4, 68.9, and 67.3 seconds per build, respectively, with a maximum value ($S_1$) above 450 seconds. This can be explained by the fact that these feature groups require static source code and dependency analysis. Conversely, TES\_CHN, with an average of 1.3 seconds per build, shows the lowest data collection time, since this feature group is based on analyzing test case source code changes in a build, and does not require any historical data or preprocessing. Further, the source code of regression test cases is not frequently changed.

Further, Table~\ref{tab:testing-vs-collection-time} shows the ratio of the total data collection time over the regression testing time for all subjects. The results suggest that data collection time can increase the regression testing time between 1\% and 71\% with an average of 11\% across subjects. As expected, larger systems and test suites tend to correspond to much larger percentages, e.g., 44\% for $S_1$, 71\% for $S_3$, and our analysis shows that the percentage of data collection time and subject size (SLOC) are strongly correlated (Spearman's $\rho$) with $\rho=0.79$. This can cause practical issues in a CI context since regression testing should be fast enough to enable the code to be built and tested several times a day. As a result, this justifies our attempt to investigate whether all feature groups are required to achieve satisfactory TCP effectiveness, especially those groups entailing the largest preprocessing times.


\textbf{EXP1.1} To answer RQ1.1, we collected data for all builds of each subject and stored the resulting datasets. We recorded the data collection time related to preprocessing and measurement for each feature group, which is used to answer RQ1.1.

\begin{table}[ht]
\caption{Results of pairwise comparison of the data collection time of feature groups using the Wilcoxon Signed-rank test and the Common Language (CL) effect size. Also, each feature group is indicated with a number that shows their rank based on average data collection time per build (higher ranks indicate higher data collection time).}
\label{tab:rq1-wilcoxon-res}
\centering
\resizebox{0.85\columnwidth}{!}{%
\begin{tabular}{ll|rr}
\toprule
\multicolumn{2}{c|}{Feature Group Pair}          &  \textit{p-value} &    CL \\
\midrule
\multirow{8}{*}{COD\_COV\_COM \circled{1}} & TES\_CHN &     0.00 &  1.00 \\
    & TES\_PRO &     0.00 &  0.98 \\
    & REC &     0.00 &  0.95 \\
    & TES\_COM &     0.00 &  0.80 \\
    & F\_COV &     0.00 &  0.60 \\
    & COD\_COV\_CHN &     0.00 &  0.59 \\
    & COD\_COV\_PRO &     0.00 &  0.59 \\
    & DET\_COV &     0.00 &  0.59 \\
\cline{1-4}
\multirow{7}{*}{COD\_COV\_PRO \circled{2}} & TES\_CHN &     0.00 &  0.99 \\
    & TES\_PRO &     0.00 &  0.95 \\
    & REC &     0.00 &  0.92 \\
    & TES\_COM &     0.00 &  0.71 \\
    & F\_COV &     0.00 &  0.51 \\
    & DET\_COV &     0.00 &  0.51 \\
    & COD\_COV\_CHN &  $<0.01$ &  0.50 \\
\cline{1-4}
\multirow{6}{*}{DET\_COV \circled{3}} & TES\_CHN &     0.00 &  0.99 \\
    & TES\_PRO &     0.00 &  0.95 \\
    & REC &     0.00 &  0.92 \\
    & TES\_COM &     0.00 &  0.71 \\
    & F\_COV &     0.00 &  0.51 \\
    & COD\_COV\_CHN &  $<0.01$ &  0.50 \\
\cline{1-4}
\multirow{5}{*}{COD\_COV\_CHN \circled{4}} & TES\_CHN &     0.00 &  0.99 \\
    & TES\_PRO &     0.00 &  0.95 \\
    & REC &     0.00 &  0.92 \\
    & TES\_COM &     0.00 &  0.71 \\
    & F\_COV &     0.00 &  0.51 \\
\cline{1-4}
\multirow{4}{*}{F\_COV \circled{5}} & TES\_CHN &     0.00 &  0.98 \\
    & TES\_PRO &     0.00 &  0.95 \\
    & REC &     0.00 &  0.92 \\
    & TES\_COM &     0.00 &  0.70 \\
\cline{1-4}
\multirow{3}{*}{TES\_COM \circled{6}} & TES\_CHN &     0.00 &  0.98 \\
    & TES\_PRO &     0.00 &  0.93 \\
    & REC &     0.00 &  0.88 \\
\cline{1-4}
\multirow{2}{*}{TES\_PRO \circled{7}} & TES\_CHN &     0.00 &  0.74 \\
    & REC &  $<0.01$ &  0.50 \\
\cline{1-4}
REC \circled{8} & TES\_CHN \circled{9} &     0.00 &  0.73 \\
\bottomrule
\end{tabular}
}
\end{table}

\textbf{RQ1.1} To check whether the differences between the average data collection times of feature groups (ref. Table~\ref{table:avg-time-subjects}) are statistically significant, we performed multiple pairwise Wilcoxon Signed-rank tests, to compare the data collection times of each feature group pair across the same builds. \TR{R1.6}{Note that the Wilcoxon Signed-rank test is a non-parametric test and does not make any distributional assumptions.}

As shown in Table~\ref{tab:rq1-wilcoxon-res} in column \textit{p-value}, the pairwise comparison indicates that the differences between all possible feature group pairs are statistically significant. As a result, we can meaningfully rank groups based on their average data collection times. The ranks are depicted in Table~\ref{tab:rq1-wilcoxon-res} with \circled{n}, where a lower rank means higher data collection time. Thus, COD\_COV\_COM has the highest data collection time, while TES\_CHN has the lowest.

We also used the Common Language (CL) effect size analysis to investigate the practical significance of differences. \TR{R1.6 R2.10}{CL is the probability of a randomly sampled item from a population being greater than a randomly sampled item from another population.} Table~\ref{tab:rq1-wilcoxon-res} shows the values of effect sizes for each pair across feature groups. The results show that feature groups that rely on coverage analysis (i.e., with COV in their name) have the smallest differences with each other and very large differences with other feature groups. \TR{2.10}{For instance, the difference of COD\_COV\_COM with other coverage-based feature groups (e.g., F\_COV) has a small effect size of around 0.6.} Additionally, except for TES\_PRO and REC, all feature pairs that do not rely on coverage analysis have large differences with each other (CL effect sizes above 0.7).

\begin{tcolorbox}
\textbf{RQ1.1}: The data collection time of feature groups that rely on coverage analysis is significantly higher than other feature groups. Feature groups that rely on test case code analysis have the second-highest, and the feature group that is based on test case execution history has the lowest data collection time.
\end{tcolorbox}


\textbf{EXP1.2} We conducted the same experiment as EXP1.1, but by only accounting for data collection time of the features for impacted files to address RQ1.2.

\begin{table}[t!]
\caption{Average data collection time for features related to impacted files per build (\textit{Avg. Impacted Collection Time}) compared to the average data collection time for all feature groups per build (\textit{Avg. Total Collection Time}) across all subjects. The times are reported in minutes, and the maximum value of each column is shown in bold.}
\label{tab:impacted-vs-collection-time}
\resizebox{\linewidth}{!}{%
\begin{tabular}{lrrr}
\toprule
$S_{ID}$ & Avg. Total Collection Time & Avg. Impacted Collection Time &  Impacted/Total (\%) \\
\midrule
 $S_{1}$ &                       \textbf{11.7} &                           \textbf{3.5} &                   29 \\
 $S_{2}$ &                        1.1 &                           0.1 &                    8 \\
 $S_{3}$ &                        5.0 &                           1.9 &                   \textbf{38} \\
 $S_{4}$ &                        1.7 &                           0.5 &                   31 \\
 $S_{5}$ &                        2.4 &                           0.8 &                   34 \\
 $S_{6}$ &                        1.4 &                           0.3 &                   25 \\
 $S_{7}$ &                        2.1 &                           0.7 &                   35 \\
 $S_{8}$ &                        0.9 &                           0.2 &                   25 \\
 $S_{9}$ &                        0.5 &                           0.1 &                   26 \\
$S_{10}$ &                        0.8 &                           0.2 &                   27 \\
$S_{11}$ &                        1.2 &                           0.4 &                   32 \\
$S_{12}$ &                        0.5 &                           0.1 &                   21 \\
$S_{13}$ &                        1.2 &                           0.1 &                    7 \\
$S_{14}$ &                        1.4 &                           0.4 &                   28 \\
$S_{15}$ &                        0.2 &                           $<0.1$ &                   10 \\
$S_{16}$ &                        0.2 &                           $<0.1$ &                   15 \\
$S_{17}$ &                        0.1 &                           $<0.1$ &                    8 \\
$S_{18}$ &                        0.6 &                           0.1 &                   22 \\
$S_{19}$ &                        0.5 &                           0.1 &                   12 \\
$S_{20}$ &                        0.7 &                           0.1 &                   14 \\
$S_{21}$ &                        0.2 &                           $<0.1$ &                    9 \\
$S_{22}$ &                        0.3 &                           $<0.1$ &                   13 \\
$S_{23}$ &                        0.2 &                           $<0.1$ &                   16 \\
$S_{24}$ &                        0.2 &                           $<0.1$ &                   20 \\
$S_{25}$ &                        0.1 &                           $<0.1$ &                   12 \\
\bottomrule
\end{tabular}
}
\end{table}

\textbf{RQ1.2} To investigate the effect of accounting for impacted files in the data collection process, we measured the data collection time of all the features which are based on impacted files. Columns \textit{Avg. Total Collection Time} and \textit{Avg. Impacted Collection Time} in Table~\ref{tab:impacted-vs-collection-time}, show the average data collection time for all features per build and the average data collection time for features based on impacted files per build, respectively, across all subjects. Also, column \textit{Impacted/Total} shows the percentage of the impacted collection time in the total collection time. The collection time of features based on impacted files takes between 7\% and 38\% of the total data collection time per build across subjects, with an average of 21\%. This is a significant portion of the data collection time and, therefore, is a strong justification to investigate whether features that are based on impacted files are required to achieve satisfactory TCP effectiveness. Our analysis also shows that such percentage has a significant correlation (Spearman's $\rho$) with both subject size (SLOC, $\rho=0.56$) and the number of test cases ($\rho=0.57$), respectively.

\begin{tcolorbox}
\textbf{RQ1.2}: On average, 21\% of the total data collection time per build is spent on collecting data for features that are based on impacted files, which is a significant portion of the collection time.
\end{tcolorbox}


\begin{figure*}[ht]
  \centering
  \caption{Linear correlation graphs between two characteristics of subjects (x-axis) and the total data collection time of all feature groups as well as two individual feature groups (y-axis).}
  \includegraphics[width=0.925\textwidth]{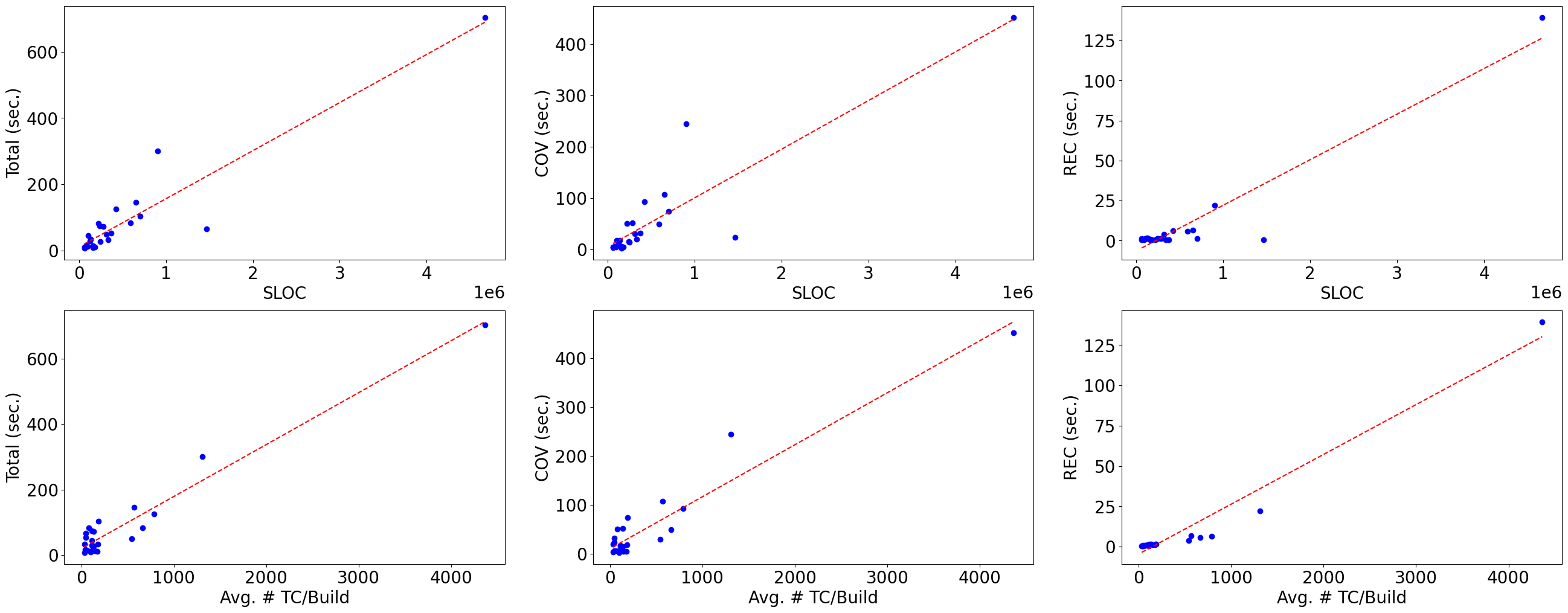}
  \label{fig:rq1-corr-graph}
\end{figure*}

\begin{table*}[ht]
\caption{\TRC{Spearman’s rank correlation between characteristics of subjects and data collection time of feature groups. Strong correlations (values greater than 0.8) are shown in bold.}}
\label{tab:rq1-corr-analysis}
\resizebox{\linewidth}{!}{%
\begin{tabular}{lrrrrrrrrrr}
\toprule
             Characteristic & COD\_COV\_COM & COD\_COV\_PRO &      DET\_COV & COD\_COV\_CHN &      F\_COV &      TES\_COM &      TES\_PRO &           REC &      TES\_CHN &         Total \\
\midrule
                 SLOC &  \textbf{0.84} &  \textbf{0.81} &  \textbf{0.81} &  \textbf{0.83} &  \textbf{0.83} &  \textbf{0.83} &     0.49 &           0.41 &     0.61 &  \textbf{0.83} \\
            Java SLOC &  \textbf{0.87} &  \textbf{0.82} &  \textbf{0.82} &  \textbf{0.85} &  \textbf{0.86} &  \textbf{0.86} &     0.46 &           0.40 &     0.57 &  \textbf{0.85} \\
           \# Commits &           0.57 &           0.57 &           0.57 &           0.60 &           0.57 &           0.51 &     0.38 &           0.60 &     0.76 &           0.63 \\
            \# Builds &           0.31 &           0.30 &           0.30 &           0.29 &           0.28 &           0.18 &     0.23 &           0.54 &     0.44 &           0.35 \\
     \# Failed Builds &           0.29 &           0.26 &           0.26 &           0.27 &           0.30 &           0.24 &    -0.04 &           0.36 &     0.31 &           0.29 \\
     Failure Rate (\%) &          -0.02 &          -0.08 &          -0.08 &          -0.05 &          -0.01 &           0.11 &    -0.29 &          -0.06 &    -0.24 &          -0.06 \\
     Avg. \# TC/Build &           0.56 &           0.56 &           0.56 &           0.56 &           0.57 &           0.40 &     0.46 &  \textbf{0.95} &     0.53 &           0.58 \\
 Avg. Test Time (min) &          -0.18 &          -0.11 &          -0.11 &          -0.15 &          -0.14 &          -0.27 &     0.12 &          -0.04 &    -0.00 &          -0.16 \\
\bottomrule
\end{tabular}
}
\end{table*}

\textbf{EXP1.3} \TR{R1.6}{To answer RQ1.3, we used Spearman’s rank correlation to assess the relationship between the data collection time of features and subject characteristics. Spearman rank correlation is a non-parametric test without any conditions about the data distribution. It is used when the variables are monotonically related and measured on a scale that is at least ordinal. The variables (i.e., data collection time of features and subject characteristics) are ordinal and in monotonic order as shown in Figure~\ref{fig:rq1-corr-graph}. Thus, it is an appropriate choice in our context.} We performed the correlation analysis between the data collection time of all features as well as each feature group, which was recorded in EXP1.1, and subject size (SLOC) as well as the number of test cases and builds.

\textbf{RQ1.3} As shown in Table~\ref{table:avg-time-subjects}, the collection time of each feature group varies across subjects. To explain such variation, we analyzed correlations between subjects' characteristics and their feature groups' collection times. \TR{R1.6}{We used Spearman’s rank correlation to assess the strength of such correlations as the inspections of scatterplots showed monotonic relationships.}

\TR{R1.6}{As shown in Table~\ref{tab:rq1-corr-analysis}, the subjects' SLOC strongly correlates with most of the feature groups' data collection times, which is not surprising since the time for coverage and code complexity analysis increases as SLOC increases. We also observe that the average number of test cases per build has a moderate correlation with the data collection time of most feature groups, and a strong correlation (0.95) with REC. REC computes features based on test execution history, and subjects with more test cases have more execution history, this results in higher collection times.} 

Figure~\ref{fig:rq1-corr-graph} depicts six strong correlations between subjects' SLOC and data collection times of the feature groups. They suggest there is an increasing monotonic relationship between SLOC and data collection time with a correlation coefficient above 0.8. We can see that there is one outlier in all the scatterplots, which refers to subject $S_1$. Thus, we repeated the correlation analysis without such outlier to investigate whether it caused significant inflation in the correlation coefficients. Though the results showed a reduction in correlation coefficients, trends remained consistent.

\begin{tcolorbox}
\textbf{RQ1.3}: The size of the subjects and the average number of test cases per build are positively and significantly correlated with the data collection time of the majority of features.
\end{tcolorbox}

\subsubsection{Training and Testing of Ranking Models for TCP (RQ2)}
\label{sec:rq2-experiments}

In the following, we assume that the builds of each subject are assigned a unique id incrementally, according to their time of occurrence.


\TR{R1.2 R1.9}{\textbf{Model Selection} Bertolino et al.~\cite{bertolinolearning} showed that Multiple Additive Regression Trees (MART)~\cite{MART}, a.k.a. Gradient boosted regression trees, is the best ML ranking model in the TCP context. Since our study uses a different evaluation metric and different subjects, we conducted an experiment to verify whether or not MART remains the best ML ranking model ($M_{opt}$) for our subjects as well and thus to obtain the best results. To this end, we evaluated MART against five other ML ranking models, namely LambdaMART (LMART)~\cite{LambdaMART}, Random Forest (RF)~\cite{random-forests}, RankBoost~\cite{RankBoost}, ListNet~\cite{ListNet}, and Coordinate Ascent (CA)~\cite{CoordinateAscent}. They were selected as they cover pairwise and listwise ML ranking models, are considered the most popular techniques in the RankLib~\cite{Ranklibs} library, and were already used by Bertolino et al.~\cite{bertolinolearning}. For all ranking models, we used an existing implementation found in the RankLib~\cite{Ranklibs} library with the default values for model hyperparameters. For each failed build with id $n$, we used all feature records from the previous failed builds for training, i.e., all feature records whose build id is between $1$ and $n-1$. We then evaluated the ranking model based on the feature records of build $n$. We conducted this experiment based on the latest 50 failed builds of each subject. For subjects with less than 50 failed builds (see Section~\ref{sec:subjects}), we used all failed builds. We then used the Friedman statistical test~\cite{friedman-test} to see if there is at least one ML model that is significantly better than others across the same builds and used the Nemenyi post hoc test~\cite{nemenyi-posthoc} to single out the best model.}

\TR{R1.2 R1.9}{The results of our comparison showed that the RF ranking model performed significantly better than the other models including MART, in contrast to the results reported by Bertolino et al.~\cite{bertolinolearning}. Therefore, we decided to use RF for all the experiments throughout the rest of this study. It is worth reminding that the main focus of this study is assessing the impact of test case features on the effectiveness of TCP. Since we expect practitioners to use the best model, we report such results for RF only. A more detailed investigation of the impact of different sets of features for different ML algorithms is out of the scope of this paper.}

\TR{R1.3}{\textbf{Hyperparameter Selection} Hyperparameter selection can significantly affect the performance of ML models. Thus, we designed an experiment to find the best hyperparameters for the RF ranking model. We investigated the following RF hyperparameters: \textit{rtype}, \textit{srate}, \textit{bag},  \textit{frate}, \textit{tree}, \textit{leaf}, and \textit{shrinkage}. Please refer to the RankLib~\cite{Ranklibs} library for details regarding these hyperparameters. We defined a hyperparameter search space by using the following values for each hyperparameter: 
$$\text{rtype}\in\{\text{MART},\text{LMART}\}, \text{srate}\in\{0.5, 1.0\},$$
$$\text{bag}\in\{150, 300, 600\}, \text{frate}\in\{0.15, 0.3, 0.6\}, \text{tree}\in\{1, 3, 5\}, $$
$$\text{leaf}\in\{50, 100, 200\}, \text{shrinkage}\in\{0.05, 0.1, 0.2\} $$
The above hyperparameter search space leads to 972 possible combinations. Since exploring all hyperparameter combinations was not computationally possible, we relied on covering arrays~\cite{cov-arr} to identify 42 combinations with the strength of 3, thus covering all hyperparameter combinations of size 3. We used all the builds (1133 builds) across subjects to evaluate the 42 combinations. Finally, we compared the results of the 42 combinations with the results of the default combination ($HC_{def}$), which was obtained from the model selection experiment, using the pairwise Wilcoxon Signed-rank test. The results showed that $HC_{def}$ was not the best but was among the best. The best combination ($HC_{opt}$) achieved an average $\text{APFD}_C$ of 0.824 that is higher than that of $HC_{def}$ with 0.813. The difference between $HC_{opt}$ and $HC_{def}$ is statistically significant, though the magnitude of the difference is small. Based on the results, the selected hyperparameter values ($HC_{opt}$) used through the rest of this study were:
$$\text{rtype}=\text{MART} , \text{srate}=0.5 , \text{bag}=150 , \text{frate}=0.3,$$
$$\text{tree}=5, \text{leaf}=200 , \text{shrinkage}=0.2$$}

\begin{tcolorbox}
\textbf{Model and Hyperparameter Selection}: The Random Forest (RF) ML ranking model performs significantly better than other investigated ranking models. In addition, the difference between the TCP effectiveness of RF's default and best hyperparameters is small though in favor of the latter.
\end{tcolorbox}


\textbf{EXP2.1} To answer RQ2.1, similar to the model selection experiment, we trained RF ranking models for the latest 50 failed builds of each subject using all feature records. Our decision for training the models only based on failed builds is based on the results of a recent study. Via empirical analysis, Elsner et al.~\cite{newbaseline} showed that training ML-based TCP models by including builds with no failures does not improve the effectiveness of the TCP models. Thus, including such builds only increases the training time without any benefit.

In practice, training ML models for each new build may not be practical or necessary, and we investigate this question in RQ3. However, since our focus is on investigating the impact of features on the effectiveness of ML-based TCP models, we train ranking models for all possible builds across subjects, to analyze results on the largest possible number of builds.

As discussed in Section~\ref{sec:metrics}, we report $\text{APFD}_C$ only based on the failed builds of all subjects. Table~\ref{tab:apfdc-results} shows $\text{APFD}_C$ averages and standard deviations of Random Forest ranking models across subjects.

\textbf{RQ2.1} Column \textit{Full\_M} in Table~\ref{tab:apfdc-results} shows $\text{APFD}_C$ averages for subjects, when the training data includes the full feature set. Average $\text{APFD}_C$ values, for all subjects except the last two, are above 0.7, and the average $\text{APFD}_C$ across all builds is 0.82. Thus, it is safe to conclude that including the full feature set in the training of ranking models for test case prioritization leads to promising results for most subjects. However, we can observe a considerable $\text{APFD}_C$ variation in Table~\ref{tab:apfdc-results} across subjects (e.g., 0.56 for $S_{25}$ and 0.98 for $S_{15}$). For this reason, we conducted a correlation analysis to measure potential correlations between $\text{APFD}_C$ and subject characteristics. As a result, we did not find any strong or moderate correlations, and therefore we left this question for future work.

\begin{tcolorbox}
\textbf{RQ2.1}: Using the full feature set for training ML ranking models for TCP leads to promising results for most subjects. However, we can observe a considerable variation in results across subjects.
\end{tcolorbox}

\begin{table}[t!]
\caption{\textbf{$\text{APFD}_C$} averages and standard deviations across subjects for four different Random Forest ranking models. \textit{Full\_M} represents the model that was trained on data that \textbf{includes} the full feature set. \textit{IMP\_M} was trained on data that \textbf{excludes} features related to impacted files. \textit{TES\_M}, \textit{REC\_M}, and \textit{COV\_M} were trained only with features related to test case source code, execution history, and test code coverage, respectively. For each subject (row), the $\text{APFD}_C$ of the model with the highest average $\text{APFD}_C$ and the lowest standard deviation is shown in bold.}
\label{tab:apfdc-results}
\resizebox{\linewidth}{!}{%
\begin{tabular}{lrrrrr}
\toprule
$S_{ID}$ &  \textit{Full\_M} &   \textit{IMP\_M} &   \textit{TES\_M} &    \textit{REC\_M} &   \textit{COV\_M} \\
\midrule
 $S_{15}$ &  $0.98 {\pm} 0.05$ &  $0.97 {\pm} 0.05$ &  $0.97 {\pm} 0.07$ &  $\mathbf{0.98 {\pm} 0.03}$ &  $0.73 {\pm} 0.24$ \\
 $S_{10}$ &  $0.94 {\pm} 0.14$ &  $0.94 {\pm} 0.14$ &  $0.92 {\pm} 0.17$ &  $\mathbf{0.94 {\pm} 0.13}$ &  $0.66 {\pm} 0.21$ \\
 $S_{14}$ &  $\mathbf{0.92 {\pm} 0.16}$ &  $0.92 {\pm} 0.17$ &  $0.92 {\pm} 0.18$ &  $0.90 {\pm} 0.20$ &  $0.60 {\pm} 0.28$ \\
  $S_{7}$ &  $0.91 {\pm} 0.17$ &  $\mathbf{0.91 {\pm} 0.16}$ &  $0.72 {\pm} 0.23$ &  $0.89 {\pm} 0.20$ &  $0.46 {\pm} 0.24$ \\
  $S_{5}$ &  $\mathbf{0.90 {\pm} 0.12}$ &  $\mathbf{0.90 {\pm} 0.12}$ &  $0.89 {\pm} 0.17$ &  $0.88 {\pm} 0.16$ &  $0.60 {\pm} 0.17$ \\
  $S_{6}$ &  $0.88 {\pm} 0.23$ &  $0.88 {\pm} 0.22$ &  $0.76 {\pm} 0.29$ &  $\mathbf{0.89 {\pm} 0.22}$ &  $0.62 {\pm} 0.29$ \\
 $S_{19}$ &  $\mathbf{0.88 {\pm} 0.12}$ &  $0.88 {\pm} 0.13$ &  $\mathbf{0.88 {\pm} 0.12}$ &  $0.87 {\pm} 0.12$ &  $0.57 {\pm} 0.25$ \\
 $S_{11}$ &  $0.86 {\pm} 0.26$ &  $0.86 {\pm} 0.26$ &  $\mathbf{0.89 {\pm} 0.21}$ &  $0.84 {\pm} 0.28$ &  $0.62 {\pm} 0.27$ \\
 $S_{23}$ &  $\mathbf{0.85 {\pm} 0.15}$ &  $\mathbf{0.85 {\pm} 0.15}$ &  $0.85 {\pm} 0.16$ &  $0.84 {\pm} 0.16$ &  $0.50 {\pm} 0.29$ \\
 $S_{12}$ &  $0.85 {\pm} 0.16$ &  $0.85 {\pm} 0.16$ &  $0.82 {\pm} 0.17$ &  $\mathbf{0.85 {\pm} 0.12}$ &  $0.62 {\pm} 0.22$ \\
  $S_{2}$ &  $0.84 {\pm} 0.04$ &  $0.84 {\pm} 0.04$ &  $\mathbf{0.85 {\pm} 0.05}$ &  $0.84 {\pm} 0.04$ &  $0.60 {\pm} 0.14$ \\
  $S_{9}$ &  $\mathbf{0.81 {\pm} 0.20}$ &  $0.81 {\pm} 0.21$ &  $0.79 {\pm} 0.24$ &  $0.80 {\pm} 0.22$ &  $0.52 {\pm} 0.12$ \\
 $S_{18}$ &  $0.81 {\pm} 0.24$ &  $0.83 {\pm} 0.22$ &  $\mathbf{0.86 {\pm} 0.17}$ &  $0.78 {\pm} 0.28$ &  $0.58 {\pm} 0.23$ \\
  $S_{4}$ &  $0.79 {\pm} 0.20$ &  $0.79 {\pm} 0.20$ &  $0.84 {\pm} 0.19$ &  $\mathbf{0.87 {\pm} 0.16}$ &  $0.53 {\pm} 0.17$ \\
  $S_{8}$ &  $\mathbf{0.78 {\pm} 0.14}$ &  $0.78 {\pm} 0.15$ &  $0.77 {\pm} 0.14$ &  $0.68 {\pm} 0.25$ &  $0.59 {\pm} 0.23$ \\
 $S_{20}$ &  $0.78 {\pm} 0.24$ &  $0.78 {\pm} 0.24$ &  $\mathbf{0.81 {\pm} 0.21}$ &  $0.78 {\pm} 0.18$ &  $0.66 {\pm} 0.26$ \\
 $S_{16}$ &  $0.78 {\pm} 0.22$ &  $\mathbf{0.79 {\pm} 0.21}$ &  $0.77 {\pm} 0.24$ &  $0.78 {\pm} 0.23$ &  $0.58 {\pm} 0.19$ \\
  $S_{1}$ &  $0.77 {\pm} 0.25$ &  $0.78 {\pm} 0.24$ &  $0.65 {\pm} 0.25$ &  $\mathbf{0.79 {\pm} 0.22}$ &  $0.42 {\pm} 0.23$ \\
 $S_{21}$ &  $0.77 {\pm} 0.23$ &  $0.77 {\pm} 0.23$ &  $\mathbf{0.80 {\pm} 0.18}$ &  $0.74 {\pm} 0.24$ &  $0.66 {\pm} 0.23$ \\
 $S_{17}$ &  $\mathbf{0.77 {\pm} 0.18}$ &  $0.76 {\pm} 0.18$ &  $0.76 {\pm} 0.18$ &  $0.73 {\pm} 0.20$ &  $0.50 {\pm} 0.12$ \\
  $S_{3}$ &  $\mathbf{0.75 {\pm} 0.25}$ &  $\mathbf{0.75 {\pm} 0.25}$ &  $0.68 {\pm} 0.24$ &  $0.74 {\pm} 0.26$ &  $0.56 {\pm} 0.21$ \\
 $S_{13}$ &  $0.73 {\pm} 0.24$ &  $0.74 {\pm} 0.24$ &  $\mathbf{0.79 {\pm} 0.23}$ &  $0.66 {\pm} 0.27$ &  $0.56 {\pm} 0.22$ \\
 $S_{24}$ &  $\mathbf{0.70 {\pm} 0.31}$ &  $0.67 {\pm} 0.31$ &  $0.69 {\pm} 0.29$ &  $0.69 {\pm} 0.31$ &  $0.60 {\pm} 0.23$ \\
 $S_{22}$ &  $0.66 {\pm} 0.29$ &  $0.67 {\pm} 0.29$ &  $\mathbf{0.71 {\pm} 0.28}$ &  $0.63 {\pm} 0.28$ &  $0.51 {\pm} 0.23$ \\
 $S_{25}$ &  $0.56 {\pm} 0.27$ &  $0.57 {\pm} 0.28$ &  $\mathbf{0.60 {\pm} 0.18}$ &  $0.54 {\pm} 0.20$ &  $0.54 {\pm} 0.09$ \\
\bottomrule
\end{tabular}
}
\end{table}


\textbf{EXP2.2} To answer RQ2.2, we conducted the same experiments as EXP2.1, but we removed features related to impacted files to compare their results with those of the all-features experiment (EXP2.1) and measure their effect on TCP effectiveness. 

\textbf{RQ2.2} Column \textit{IMP\_M} in Table~\ref{tab:apfdc-results} reports average $\text{APFD}_C$ values for subjects when features related to impacted files are not included in the training data. As shown, the differences between the $\text{APFD}_C$ of \textit{Full\_M} and \textit{IMP\_M} are small across all subjects, and they range between 0.0 and 0.03.

We also ran a Wilcoxon Signed-rank test based on $\text{APFD}_C$ results for \textit{Full\_M} and \textit{IMP\_M} across all builds and subjects. 
The test results show that there is no statistically significant difference between the $\text{APFD}_C$ of \textit{Full\_M} and \textit{IMP\_M}; \textit{p-value}=0.14. Hence, it is safe to conclude that accounting for the features of impacted files does not bring practical advantages as it does not have a significant effect, across all subjects, on the effectiveness of the ranking models in terms of $\text{APFD}_C$.

\begin{tcolorbox}
\textbf{RQ2.2}: Test case features that are related to impacted files do not have a significant effect on the effectiveness of TCP.
\end{tcolorbox}


\textbf{EXP2.3} To answer RQ2.3, we conducted two sets of experiments:
\begin{itemize}
    \item \textbf{EXP2.3.1} We trained nine ranking models for each failed build using the same method as in EXP2.1 and EXP2.2. The nine models differ based on the feature groups used for their training since for each model, one feature group was left out. 
    \item \textbf{EXP2.3.2} We trained three ranking models for each failed build using the same method as in EXP2.1 and EXP2.2. Again, the three models differ based on the features used for their training. The first model (\textit{COV\_M}) was trained using all coverage-related features, all of which are collected via source code coverage analysis of the system under test, i.e,  F\_COV, COD\_COV\_COM, COD\_COV\_PRO, COD\_COV\_CHN, and DET\_COV. The second model (\textit{TES\_M}) was trained using features that are calculated via static analysis of test case source code, i.e., TES\_COM, TES\_PRO, and TES\_CHN. The last model (\textit{REC\_M}) was trained only based on test case execution history, i.e., REC. 
    The data collection time for features in the same high-level group is heavily driven by data preprocessing of common data sources, and therefore, removing or adding features in such a group does not make a practical difference. Therefore, understanding the impact of each high-level group can help practitioners select the best feature group(s) when the collection of all features is not possible. 
\end{itemize}

As we explained in EXP2.3.2 above, the feature extraction cost of a single feature (group) within a high-level group is nearly the same as the cost of extracting all features in that group. As a complementary lower-level analysis, that is admittedly of less practical importance, we investigated the impact of removing one (low-level) feature group and the importance of each individual features in RQ2.4. Given the large volume of our data and the high execution time of the experiments when training Random Forest models, we could not train these models using all possible feature (group) combinations.

\begin{table}[t!]
\caption{Pairwise \textit{$\text{APFD}_C$} comparison results across subjects between models that exclude individual feature groups and the model that includes the full feature set (\textit{Full}). The comparison is done using the Wilcoxon Signed-rank test and the Common Language (CL) effect size. \textit{p-values} below 0.05 are shown in bold.}
\label{tab:rq2-exclude-tests}
\centering
\begin{tabular}{ll|rr}
\toprule
\multicolumn{2}{l|}{Experiment Pair} & \textit{p-value} & CL \\
\midrule
 Full &       TES\_COM &  $\mathbf{<0.01}$ &  0.51 \\
 Full &            REC &  $\mathbf{<0.01}$ &  0.54 \\
 Full &       TES\_PRO &     \textbf{0.01} &  0.51 \\
 Full &  COD\_COV\_COM &     0.13 &  0.50 \\
 Full &  COD\_COV\_CHN &     0.18 &  0.50 \\
 Full &       DET\_COV &     0.41 &  0.50 \\
 Full &  COD\_COV\_PRO &     0.52 &  0.50 \\
 Full &         F\_COV &     0.56 &  0.50 \\
 Full &       TES\_CHN &     0.97 &  0.50 \\
\bottomrule
\end{tabular}
\end{table}

\begin{table}[t!]
\caption{Pairwise \textit{$\text{APFD}_C$} comparison results across subjects between models that include individual high-level feature groups and the model that includes the full feature set (\textit{Full}). The comparison is done using the Wilcoxon Signed-rank test and the Common Language (CL) effect size. \textit{p-values} below 0.05 are shown in bold.}
\label{tab:rq2-include-tests}
\centering
\begin{tabular}{ll|rr}
\toprule
\multicolumn{2}{l|}{Experiment Pair} & \textit{p-value} & CL \\
\midrule
 Full &  COV\_M &  $\mathbf{<0.01}$ &  0.79 \\
 Full &  REC\_M &  $\mathbf{<0.01}$ &  0.51 \\
 Full &  TES\_M &  $\mathbf{<0.01}$ &  0.53 \\
\bottomrule
\end{tabular}
\end{table}

\textbf{RQ2.3} The results of EXP2.3.1 show that excluding any one of the feature groups does not cause a significant decrease in effectiveness for the trained models. We conducted nine Wilcoxon Signed-rank tests on $\text{APFD}_C$ results for all builds across subjects. As shown in Table~\ref{tab:rq2-exclude-tests}, among all nine feature groups, removing TES\_COM, REC, or TES\_PRO causes a statistically significant difference in $\text{APFD}_C$ with \textit{Full\_M}; TES\_COM: \textit{p-value}\textless0.01, REC: \textit{p-value}\textless0.01, and TES\_PRO: \textit{p-value}=0.01. However, the CL effect size analysis shows that the magnitude of their impact is practically negligible (CL is around 0.5). Based on the above results, we can conclude that removing any of the nine feature groups does not have a practical impact on the $\text{APFD}_C$ of the ranking models.

Columns \textit{TES\_M}, \textit{REC\_M}, and \textit{COV\_M} in Table~\ref{tab:apfdc-results} report the average $\text{APFD}_C$ of ranking models trained with features related to test case source code, execution history, and test code coverage, respectively. Though the $\text{APFD}_C$ of \textit{COV\_M} is low, \textit{REC\_M} reaches similar $\text{APFD}_C$ to \textit{Full\_M}. To analyze the result of EXP2.3.2, we conducted three Wilcoxon Signed-rank tests to assess the differences of the three models with \textit{Full\_M}. As shown in Table~\ref{tab:rq2-include-tests}, the $\text{APFD}_C$ results of all three models are statistically different from \textit{Full\_M}; \textit{p-value}$\leq$0.01. The CL effect size analysis for $\text{APFD}_C$ results shows that the difference with \textit{Full\_M} for \textit{COV\_M}, with a value of 0.79, is practically significant, as opposed to the CL effect sizes of \textit{TES\_M} and \textit{REC\_M}, which are 0.53 and 0.51, respectively. Thus we can conclude that, among all high-level feature groups, \textit{TES\_M}---which is trained with features related to the source code of test cases--- and \textit{REC\_M}---which is trained with features related to the execution history of test cases--- achieve the best results, with nearly the same $\text{APFD}_C$ as the models trained with the full feature set. Further, the models trained with only the coverage-based feature groups achieve the worst $\text{APFD}_C$ and are not reliable individually.

\begin{tcolorbox}
\textbf{RQ2.3}: Each of the nine feature groups does not have a significant impact on TCP effectiveness, individually. This is different for high-level feature groups. Features related to test case source code or execution history have nearly the same impact on TCP effectiveness as all features, whereas coverage-based features have a significantly lower impact.
\end{tcolorbox}


\textbf{EXP2.4} To address RQ2.4, we rely on RankLib~\cite{Ranklibs} as it provides model feature statistics for the RF model. It captures how frequently each feature is used to split nodes of trained regression trees. Such decision trees use information gain, which is based on the concept of entropy, to determine which feature to use to further split the tree at each training step. Therefore, a higher usage frequency in the tree for a feature indicates that it tended to yield higher information gains at training time. We use feature usage frequencies to measure the impact of individual features in our feature groups.

\begin{table}[t!]
\caption{Top 15 frequently used individual features in models that were trained on the full feature set. The \textit{Avg. Freq.} column shows the average usage frequency of a feature across all ranking models.}
\label{tab:rq2-fstats-top15}
\centering
\resizebox{0.85\linewidth}{!}{%
\begin{tabular}{llr}
\toprule
Feature Group &                Feature &  Avg. Freq. \\
\midrule
          REC &                     Age &       17034 \\
     TES\_PRO &        OwnersExperience &       10618 \\
     TES\_PRO &  AllCommitersExperience &        7643 \\
          REC &         TotalMaxExeTime &        5409 \\
          REC &             LastExeTime &        4095 \\
     TES\_PRO &      OwnersContribution &        3997 \\
     TES\_PRO &             CommitCount &        3986 \\
          REC &         TotalAvgExeTime &        3977 \\
          REC &        RecentAvgExeTime &        3850 \\
          REC &        RecentMaxExeTime &        3741 \\
     TES\_COM &      RatioCommentToCode &        3695 \\
     TES\_COM &           CountStmtDecl &        3383 \\
     TES\_COM &       CountLineCodeDecl &        3342 \\
     TES\_COM &          CountLineBlank &        3149 \\
     TES\_COM &            CountStmtExe &        2825 \\
\bottomrule
\end{tabular}
}
\end{table}

\begin{figure}[t!]
  \caption{The histogram shows the relationship between the age of all test cases and their failure frequency. The x-axis shows the normalized age between 0 and 100 and the y-axis shows the number of failures of all test cases in a given age period.}
  \includegraphics[width=0.9\linewidth]{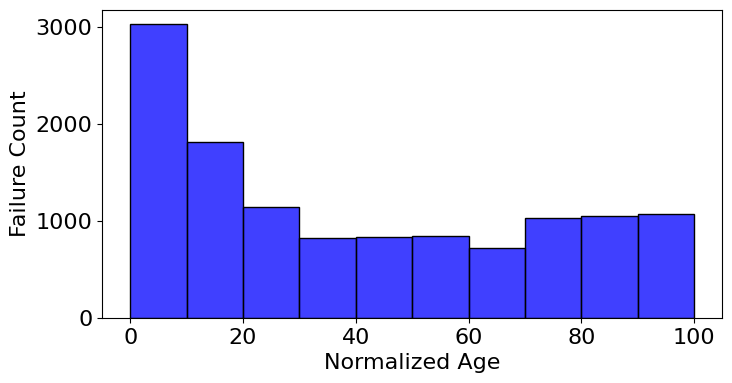}
  \label{fig:rq2-age-hist}
\end{figure}

\textbf{RQ2.4} Table~\ref{tab:rq2-fstats-top15} reports the average usage frequency of the top 15 features in \textit{Full\_M} for models across all subjects. Top features belong to the three feature groups REC, TES\_PRO, and TES\_COM, all of which are identified as significant feature groups by RQ2.3.

From the TES\_COM group, features capturing the size of test cases show higher usage, which is to be expected as large test cases tend to have a higher execution time and coverage. Also, from the TES\_PRO group, features capturing the experience of test case developers and the number of commits are more important. We conjecture that experienced developers may better understand the system under test to develop high-quality test cases compared to less experienced developers. Also, higher numbers of commits for a test case suggest that the test case is under active development and is being updated to detect faults.

From the REC feature group, features related to previous execution times and the age of test cases are the most frequently used. The former are good indicators of future test case execution times and explain why execution time-related features have high usage. However, seeing the age of a test case as the most important among all features may seem surprising at first glance. To explain this, we analyzed the relation between the age of test cases and their failure occurrences across all subjects. As shown in Figure~\ref{fig:rq2-age-hist}, failures drop sharply after the first 10\% and 20\% of the builds on which they are executed. Thus, the age of a test case is a good predictor of its failure probability. Further, this suggests that all subjects contain obsolete test cases that are not actively maintained and not kept updated. When a new functionality of the SUT is developed, a number of test cases are developed to test it, and once the functionality works as expected and becomes stable over time, their corresponding test cases often pass across builds.

\begin{tcolorbox}
\textbf{RQ2.4}: The most frequently used test case features in the ML ranking models belong to REC, TES\_PRO, and TES\_COM feature groups. These features include test case size, development history, failure history, and age.
\end{tcolorbox}

\TRC{\textbf{EXP2.5} A heuristic-based TCP approach prioritizes test cases based on single features of test cases~\cite{newbaseline}. Given a test suite $T$ and a test case feature $f_i$, we define two heuristic-based ranking models $M_{f_i, a}$ and $M_{f_i, d}$. $M_{f_i, a}$ produces a test case ordering $T^{\ast}_a$ in which test cases are sorted by their corresponding value of $f_i$ in ascending order. Conversely, $M_{f_i, d}$ produces $T^{\ast}_d$ in which test cases are sorted in descending order.}

\TR{R1.4 R2.12}{Elsner et al.~\cite{newbaseline} reported that heuristic-based TCP models that are based on \textit{F\_MaxTestFileFailRate} (defined in \ref{rec-def}) outperform other heuristics. We first investigated if the same results held with the subjects of this study. To do that, for each test case feature $f_i$ ($1 \leq i \leq 150$), we applied the $M_{f_i, a}$ and $M_{f_i, d}$ heuristic-based models (300 models in total) and compared their results. We observed that the $M_{f_{56}, d}$ model, which is based on the \textit{F\_FailRate(Total)} feature (Section~\ref{rec-def}), performed best for the majority of subjects, in contrast to previous work~\cite{newbaseline}. Elsner et al.~\cite{newbaseline} also concluded that heuristic-based TCP models are better than ML-based TCP models in terms of average $\text{APFD}_C$, though not statistically different. Thus, in this experiment, we compared the results of our ML-based TCP model, RF, with the best heuristic, i.e., \textit{F\_FailRate(Total)}.}

\textbf{RQ2.5} To compare the effectiveness of the best \textit{H\_M} (heuristic-based) model with the \textit{Full\_M} ML-based model, we conducted a Wilcoxon Signed-rank test for all builds across subjects. The test results showed that the difference between the two models is statistically significant (\textit{p-value}$<0.01$), and the \textit{Full\_M} model, with an average $\text{APFD}_C$ of 0.82, clearly outperforms the \textit{H\_M} model, with an average $\text{APFD}_C$ of 0.71. 

\begin{table}[ht]
\caption{Pairwise $\text{APFD}_C$ comparison across subjects between the heuristic model (\textit{H\_M}) based on \textit{F\_FailRate Total} and ML-based models that were trained on the full feature set (\textit{Full\_M}). The comparison is done using the Wilcoxon Signed-rank test and the Common Language (CL) effect size. For each subject (row), the $\text{APFD}_C$ of the model with the higher average $\text{APFD}_C$ and the lower standard deviation is shown in bold. Also, \textit{p-values} below 0.05 are shown in bold.}
\label{tab:rq2-apfdc-heurisitc}
\centering
\resizebox{0.75\linewidth}{!}{
\begin{tabular}{l|rrrr}
\toprule
$S_{ID}$ &  \textit{Full\_M} &     \textit{H\_M} & \textit{p-value} &   CL \\
\midrule
  $S_{2}$ &  $\mathbf{0.84 {\pm} 0.04}$ &  $0.82 {\pm} 0.05$ &  $\mathbf{<0.01}$ &  0.86 \\
 $S_{16}$ &  $\mathbf{0.78 {\pm} 0.22}$ &  $0.58 {\pm} 0.16$ &  $\mathbf{<0.01}$ &  0.78 \\
 $S_{19}$ &  $\mathbf{0.88 {\pm} 0.12}$ &  $0.50 {\pm} 0.38$ &  $\mathbf{<0.01}$ &  0.74 \\
 $S_{21}$ &  $\mathbf{0.77 {\pm} 0.23}$ &  $0.49 {\pm} 0.33$ &  $\mathbf{<0.01}$ &  0.73 \\
  $S_{6}$ &  $\mathbf{0.88 {\pm} 0.23}$ &  $0.68 {\pm} 0.30$ &  $\mathbf{<0.01}$ &  0.72 \\
 $S_{18}$ &  $\mathbf{0.81 {\pm} 0.24}$ &  $0.60 {\pm} 0.33$ &  $\mathbf{<0.01}$ &  0.69 \\
 $S_{20}$ &  $\mathbf{0.78 {\pm} 0.24}$ &  $0.57 {\pm} 0.32$ &  $\mathbf{<0.01}$ &  0.68 \\
 $S_{15}$ &  $\mathbf{0.98 {\pm} 0.05}$ &  $0.86 {\pm} 0.30$ &  $\mathbf{<0.01}$ &  0.66 \\
  $S_{1}$ &  $\mathbf{0.77 {\pm} 0.25}$ &  $0.58 {\pm} 0.32$ &  $\mathbf{<0.01}$ &  0.66 \\
  $S_{7}$ &  $\mathbf{0.91 {\pm} 0.17}$ &  $0.84 {\pm} 0.25$ &     0.13 &  0.65 \\
  $S_{3}$ &  $\mathbf{0.75 {\pm} 0.25}$ &  $0.54 {\pm} 0.34$ &  $\mathbf{<0.01}$ &  0.64 \\
 $S_{17}$ &  $\mathbf{0.77 {\pm} 0.18}$ &  $0.69 {\pm} 0.23$ &     0.07 &  0.62 \\
  $S_{9}$ &  $0.81 {\pm} 0.20$ &  $\mathbf{0.83 {\pm} 0.12}$ &     0.30 &  0.62 \\
 $S_{13}$ &  $\mathbf{0.73 {\pm} 0.24}$ &  $0.64 {\pm} 0.29$ &     0.43 &  0.61 \\
 $S_{12}$ &  $\mathbf{0.85 {\pm} 0.16}$ &  $0.77 {\pm} 0.21$ &     \textbf{0.02} &  0.60 \\
  $S_{5}$ &  $\mathbf{0.90 {\pm} 0.12}$ &  $0.74 {\pm} 0.34$ &     \textbf{0.01} &  0.59 \\
 $S_{22}$ &  $\mathbf{0.66 {\pm} 0.29}$ &  $0.55 {\pm} 0.35$ &     0.08 &  0.58 \\
  $S_{8}$ &  $\mathbf{0.78 {\pm} 0.14}$ &  $0.71 {\pm} 0.18$ &     0.07 &  0.56 \\
 $S_{14}$ &  $\mathbf{0.92 {\pm} 0.16}$ &  $0.84 {\pm} 0.28$ &  $\mathbf{<0.01}$ &  0.56 \\
 $S_{11}$ &  $0.86 {\pm} 0.26$ &  $\mathbf{0.86 {\pm} 0.21}$ &     0.39 &  0.55 \\
 $S_{24}$ &  $\mathbf{0.70 {\pm} 0.31}$ &  $0.66 {\pm} 0.32$ &     0.44 &  0.54 \\
 $S_{25}$ &  $0.56 {\pm} 0.27$ &  $\mathbf{0.59 {\pm} 0.28}$ &     0.81 &  0.48 \\
 $S_{10}$ &  $\mathbf{0.94 {\pm} 0.14}$ &  $0.94 {\pm} 0.22$ &     0.53 &  0.46 \\
 $S_{23}$ &  $\mathbf{0.85 {\pm} 0.15}$ &  $0.85 {\pm} 0.22$ &     0.27 &  0.45 \\
  $S_{4}$ &  $0.79 {\pm} 0.20$ &  $\mathbf{0.89 {\pm} 0.16}$ &     \textbf{0.04} &  0.42 \\
\bottomrule
\end{tabular}
}
\end{table}

For further analysis, we compared \textit{H\_M} and \textit{Full\_M} by conducting Wilcoxon Signed-rank tests for each subject. Table~\ref{tab:rq2-apfdc-heurisitc} shows the statistical test results as well as $\text{APFD}_C$ averages and standard deviations across subjects. Based on Table~\ref{tab:rq2-apfdc-heurisitc}, \textit{H\_M} outperformed \textit{Full\_M} in only four subjects in terms of average $\text{APFD}_C$, and in only one of these four subjects, the difference was statistically significant. For the rest of the subjects, \textit{Full\_M} achieved a higher average $\text{APFD}_C$, and in 13 subjects, the difference was statistically significant.

Further, using Spearman's rank correlation coefficient ($\rho$), we investigated the relationship between subject characteristics and their \textit{Full\_M} and \textit{H\_M} average $\text{APFD}_C$. The analysis showed that a moderate correlation, with $\rho=-0.28$, exists between subject failure rate and the difference between \textit{Full\_M} and \textit{H\_M} in terms of average $\text{APFD}_C$. Thus, the \textit{H\_M} model tends to perform better than \textit{Full\_M} in subjects that have higher failure rates, especially in the presence of a large number of failed builds (e.g., $S_4$), which can be explained since \textit{H\_M} is based on the failure rate heuristic. However, Beller et al.~\cite{beller2017oops} conducted a comprehensive analysis of TravisCI projects and showed that for all 1,108 Java projects with test executions, the ratio of builds with at least one failed test case has a median of 2.9\% and a mean of 10.3\%. Therefore, since high failure rates (e.g., failure rates above 30\% in our subjects) are not common in CI contexts, \textit{H\_M} is probably not a good option in practice.

To conclude, our empirical analysis shows that the Random Forest TCP ranking model, when trained on the full feature set, performs overall significantly better than the best heuristic-based model based on the \textit{F\_FailRate (Total)} feature. Nevertheless, the heuristic-based model performs significantly better for a few subjects. Though this remains to be confirmed by further studies, an analysis of the characteristics of these projects suggests this is due to a combination of large numbers of failed builds with high failure rates, which are unlikely in a CI context.

\begin{tcolorbox}
\textbf{RQ2.5}: The Random Forest ranking model, when trained on the full feature set, significantly outperforms the best heuristic-based model for most subjects. However, the heuristic-based model performs significantly better for a few subjects, all of which have high build failure rates.
\end{tcolorbox}
\subsubsection{Effectiveness Decay of ML-Based TCP Models (RQ3)}

\TR{R1.7}{\textbf{EXP3} To answer RQ3, we used all of the pretrained models from EXP2.1 and tested each model on the subsequent builds using features that were computed based on preprocessed data (i.e., dependency graph, source code, and process metrics) of the build related to the model. Assume we have $n$ builds ($B = \{b_1 , b_2 , \dotsm , b_n \}$) and $n$ pretrained models ($M = \{m_1 , m_2 , \dotsm , m_n \}$), and the pretrained model of $b_i$ is $m_i$ where $1 \leq i \leq n$. In this experiment, we used $b_k$ and $m_k$ ($1 \leq k \leq n$ and $n = 50$ for the latest 50 failed builds) from EXP2.1 and all its relevant preprocessed data. Then, for each $b_i$ after $b_k$, i.e., $\{b_{k+1} , b_{k+2} , \dotsm , b_n\}$, we recomputed its test case features based on the preprocessed data available in $b_k$, ranked its test cases using $m_k$, and evaluated the ranking. The main motivation for reusing pretrained ranking models is to reduce the cost of data preprocessing and model training, and for this reason, we wanted to assess the impact of such reuse on $\text{APFD}_C$ decay. When the preprocessed data of $b_k$ did not contain enough information to compute some of the test case features in $b_i$ (i.e., new test cases or new code coverage graphs), we used the mean substitution imputation method for missing values, as it is a simple, efficient, and effective method~\cite{meanSubstitution}, and assigned those missing features with the average value across all other test cases with such feature values.}

\TR{R1.7}{\textbf{Results} As we discussed in \textit{EXP3}, we have tested the pretrained ranking model for each failed build on its following failed builds to measure how the $\text{APFD}_C$ of the pretrained models decreases over time. Here we assume a retraining window (RW) as the distance (i.e., number of builds) between the builds based on which the model is trained and tested, respectively, e.g., when $m_k$, that is trained based on the data captured up to $b_k$, is used for TCP in builds up to $b_{k+4}$ without retraining, then $\text{RW}=4$. Table~\ref{tab:rw-example} shows an example of RWs when the number of builds is equal to 5. Each cell with row $b_i$ and column $m_j$ in Table~\ref{tab:rw-example} is equal to the RW when the pretrained model $m_j$ is used to prioritize the test cases of $b_i$.}

\begin{figure}[t!]
  \caption{This graph shows the relationship between the retraining window (RW) of ranking models across all subjects and their average $APFD_C$.}
  \includegraphics[width=\linewidth]{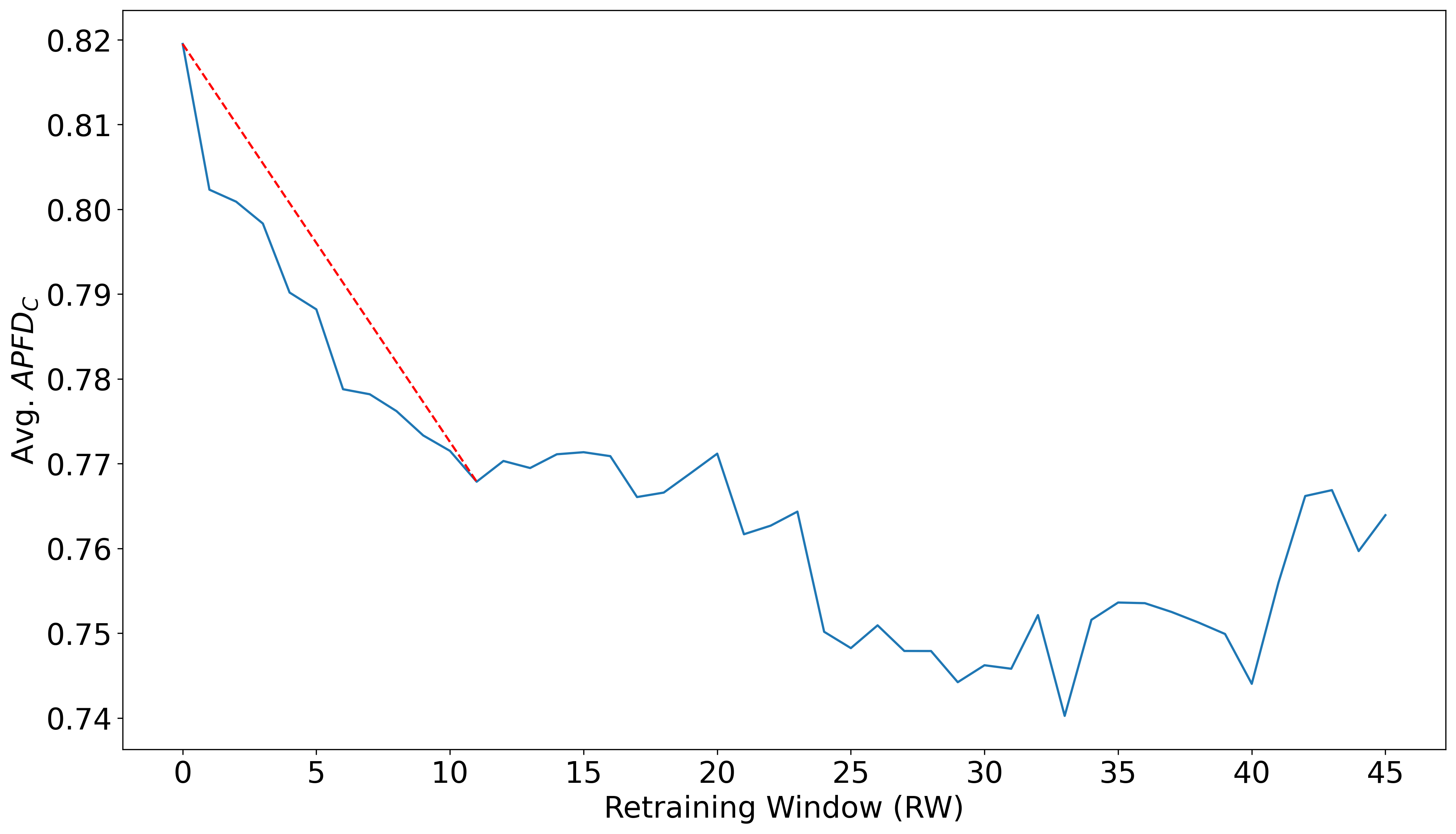}
  \label{fig:rq3-apfdc-decay}
\end{figure}

\TR{R1.7}{We computed the $\text{APFD}_C$ for all valid builds and pretrained models across subjects. Figure~\ref{fig:rq3-apfdc-decay} shows the $\text{APFD}_C$ of pretrained models, based on all builds across subjects, for RWs ranging from 0 to 45. The x-axis is the RW, and the y-axis is the average $\text{APFD}_C$ of all pretrained models with the same RW. As shown in Figure~\ref{fig:rq3-apfdc-decay}, there is a steady decreasing $\text{APFD}_C$ trend until RW is about 11. The two points $\text{RW}=0$ and $\text{RW}=11$ are connected with a red dotted line in Figure~\ref{fig:rq3-apfdc-decay}. The slope of this line, which is the average decrease in $\text{APFD}_C$ after each RW increment, is equal to -0.005. However, $\text{APFD}_C$ values above $\text{RW}=11$ display unstable trends. Hence, we can conclude that retraining ranking models with a RW of less than 11 builds is necessary to obtain predictable results across all subjects. However, given the available budget for data collection and model retraining, we also conclude that ML ranking models should be retrained as frequently as possible to achieve the best possible TCP effectiveness.}

\begin{table}[t!]
\caption{An example of retraining window (RW) values when the number of builds is equal to 5. Each cell shows the RW when TCP is applied to the build of the cell's row using the pretrained model of the cell's column. For instance, if we use $m_1$ for $b_4$, then $RW = 3$.}
\label{tab:rw-example}
\centering
\resizebox{0.55\linewidth}{!}{%
\begin{tabular}{|l|r|r|r|r|r|}
\hline
      &$m_1$& $m_2$& $m_3$& $m_4$& $m_5$ \\ \hline
$b_1$ & 0   & -    & -    & -    & -    \\ \hline
$b_2$ & 1   & 0    & -    & -    & -    \\ \hline
$b_3$ & 2   & 1    & 0    & -    & -    \\ \hline
$b_4$ & 3   & 2    & 1    & 0    & -    \\ \hline
$b_5$ & 4   & 3    & 2    & 1    & 0   \\ \hline
\end{tabular}%
}
\end{table}

\begin{tcolorbox}
\textbf{RQ3}: Retraining ranking models with a retraining window (RW) of less than 11 builds is necessary to obtain predictable results. Also, the RW should be as small as possible to obtain the best TCP effectiveness.
\end{tcolorbox}

\subsubsection{Trade-off Between Data Collection Time and TCP Effectiveness for Features (RQ4)}
Based on the results of RQ1-3, three different decisions can be taken concerning the application of TCP techniques for a software system to achieve the best trade-off, in a given context, between data collection overhead and TCP effectiveness. In the following, we provide guidelines regarding such decisions.

\textbf{Use ML model relying on the full feature set with retraining at each build or every 11 builds.} When higher TCP effectiveness is essential, and the cost of data collection can be afforded, the use of the full feature set for training ML-based TCP models is recommended since, as shown in RQ2, it yields the best results. If necessary, based on the results of RQ3, to decrease data collection time at the cost of a slight effectiveness loss, model retraining can be conducted every 11 builds instead of each build. 

\textbf{Use ML model relying on the REC feature set.} When the overhead of the full feature set cannot be afforded, regardless of the training strategy, and when a small decrease in model effectiveness is acceptable given a much lower data collection time, an ML model relying on the REC feature set is recommended. As discussed in RQ1, the data collection time of the REC feature group (i.e., features based on test execution history) is in the order of seconds and negligible. Further, RQ2 results show that not only removing the REC feature group causes statistically significant differences in model effectiveness, but furthermore training the model with only the REC features achieves results that are close to those obtained with the full feature set. Finally, our comparison between ML models based on REC features and heuristics models shows that the former's effectiveness is higher than the latter's. 

\textbf{Use failure-based heuristics.} When no data collection overhead is acceptable, the use of failure-based heuristics can be effective, especially in the context of systems with a high number of builds and high failure rates, thus potentially enabling the definition of effective heuristics. 

\begin{tcolorbox}
\textbf{RQ4}: To achieve the best TCP effectiveness when data collection cost is affordable, we recommend using Random Forest (RF) ranking models that are trained on the full feature set. When the data collection is not affordable, we suggest an RF model relying on the REC feature set. Finally, the fastest and least effective approach is the use of failure-based heuristics.
\end{tcolorbox}

\subsubsection{Configuration}
Data collection for each subject was conducted using one CPU core (Intel Xeon Gold 6234 CPU @ 3.30GHz) with a memory usage ranging between 1GB to 60GB, depending on the size of the subjects. Also, our data collection tool, all results, and datasets are made publicly available\footnote{\underline{\url{https://github.com/Ahmadreza-SY/TCP-CI}}}. We used PyDriller~\cite{PyDriller} for mining git repositories, Understand~\cite{understand} for static and dependency analysis of source code, and RankLib~\cite{Ranklibs} for training our TCP machine learning ranking models.

\subsection{Threats to Validity}

\subsubsection{Internal Threats}
\TR{R2.2}{As explained in Section~\ref{sec:dependencyanalysis}, for scalability reasons invoked by the industry partner supporting this research, the test case code coverage that we computed is an estimation based on static code coverage, which is computed using Understand~\cite{understand}, and the history of file changes in the system’s code repository. This, however, can lead to overestimating test case code coverage. Also, based on Section~\ref{sec:feature-model}, our proposed test case coverage is at the file level, which can further worsen the overestimation issue. Future work should systematically investigate trade-offs between finer-grained coverage measurement, scalability, and TCP effectiveness.}

In Travis CI, each CI cycle (build) includes one or more jobs, each of which tests the source code for different configurations and platforms (e.g., a project may have two jobs to test the code for Java 1.6 and 1.8). In most cases, the jobs share the same test cases (98\% of test cases for 24 out of 25 of our subjects are the same across jobs of the same build), resulting in more than one execution record for the repeated test cases that affect the quality of the datasets negatively. To mitigate this issue, we only focused on a job with the highest number of test cases. This reduces our data sets size in terms of the number of test cases around 2\% that may have a negligible effect on our experiments since the large number of builds that are used in the experiments compensate for this reduction. An alternative way to deal with this issue is to include extra features concerning the platform. However, we left this to future work. 

Also, as discussed in Section~\ref{sec:rq2-experiments}, we removed frequent-failing test cases using the three-sigma rule technique that may deflate the results of the ML models. To investigate this, we measured the difference between the effectiveness of our ML models with/without outlier test cases that showed there is no statistically significant difference between the two cases, i.e., removing outlier test cases does not deflate the results of the ML models.

\subsubsection{External Threats}
We relied on \textit{GHTorrent}~\cite{ghtorrent} to search for open-source projects. In addition, \textit{TravisTorrent}~\cite{travisTorrent} was used to fetch build logs from Travis CI and to find the commits that are related to builds. We also included a subset of test case execution record datasets that were published by \textit{RTPTorrent}~\cite{RTPTorrent}. There might be flaws in these tools and datasets---though it is unlikely given how widely used they are---that impact the quality of our data collection.

\section{Related work}
\label{sec:related-work}
Many studies have addressed Test Case Prioritization (TCP) in the regression testing context. Overall, these studies can be classified into two groups: heuristic-based and ML-based techniques. Since our work is ML-based, in the following, we only briefly discuss the heuristic-based group.

\textbf{Heuristic-based TCP.} The studies in this group apply heuristics to TCP that are mainly defined based on test coverage and execution history. 

Coverage-based heuristics are based on the principle that increasing structural test coverage of the System Under Test (SUT) increases the chances of fault detection~\cite{kim2002history}. Structural coverage measures include statement coverage \cite{rothermel1999test}, functions/methods coverage~\cite{rothermel2001prioritizing}, and modified condition/decision coverage  \cite{jones2003test}. Overall, coverage-based heuristics can be grouped into two sub-groups: total coverage and additional coverage~\cite{jeffrey2006test}. At each prioritization step, the former selects the test case with the highest coverage, while the latter selects the test case with the highest coverage of entities not yet covered by higher-priority test cases. 

Extracting precise coverage information at a reasonable cost is the major drawback of coverage-based heuristics.  Coverage information can be collected either by static or dynamic analysis. Static analysis techniques are more easily amenable to collecting coverage and conducting impact analysis. However, they overestimate the coverage and impact of test cases, and relying solely on them may undermine the effectiveness of TCP. Dynamic analysis is the most precise technique and requires the execution of tests on the instrumented (i.e., binary or source code instrumentation) version of the SUT. Dynamic analysis techniques are platform-dependent, time-consuming for large codebases, and may not be applicable for real-time systems because code instrumentation may cause timeouts or interrupt normal test execution. Overall, dynamic analysis techniques are difficult or even impossible to apply in practice, specifically in the CI context, mainly due to computational overhead and applicability~\cite{elbaum2014techniques,spieker2017reinforcement,lima2020learning,do2020multi,memon2017taming}. Thus, our work only investigates the use of static coverage information. We conjecture that mixing static coverage information with other relevant information, such as test cases' execution history, can compensate for the static analysis techniques' tendency to overestimate.

Heuristics that are based on test case execution history assume that previously failed test cases are more likely to fail again. For example, Kim and Porter \cite{kim2002history} proposed an execution history heuristic that calculates ranking scores based on the average failure rate of test cases. The execution history is easily accessible, and some studies such as ~\cite{spieker2017reinforcement,newbaseline} reported that the effectiveness of execution history heuristics is almost the same as ML-based TCP. However, as discussed by Pan et al.~\cite{oursurvey}, the reported training of ML models is problematic due to (1) limited feature sets, (2) unbalanced datasets containing only a small number of failed test cases, and (3) suboptimal ML techniques in the TCP context. To further investigate, we performed a comparison between the best heuristics based on the failure history of test cases and the best ML models, as discussed in Section~\ref{sec:validation}.

\textbf{ML-based TCP:} Work in this group trains ML models based on features collected from different sources, such as test execution history, to prioritize test cases. Compared to heuristics that are often defined based on a single feature, ML models can be trained on a set of features from different sources. This enables ML models to deal with the complexities of the TCP problem for complex systems in the CI context. Also, ML techniques either support online training or can be retrained to account for the dynamic nature of CI~\cite{facebookTSP}. However, heuristics are static, and there is no standard procedure to tune them based on new changes.

Pan et. al~\cite{oursurvey} surveyed ML-based TCP techniques and reported that existing ML-based papers investigated a wide variety of ML techniques including reinforcement learning~\cite{spieker2017reinforcement,bertolinolearning,do2020multi,lima2020learning,lima2020multi,rosenbauer2020xcs,shi2020reinforcement}, clustering~\cite{almaghairbe2017separating,carlson2011clustering,chen2011using,kandil2017cluster,khalid2019weight,wang2012using,yoo2009clustering}, ranking models~\cite{bertolinolearning,busjaeger2016learning,chen2018optimizing,hasnain2019recurrent,jahan2019version,lachmann2016system,mahdieh2020incorporating,mirarab2008empirical,noor2017studying,palma2018improvement,article,singhmachine,tonella2006using}, and natural
language processing~\cite{kandil2017cluster,busjaeger2016learning,lachmann2016system,aman2020comparative,thomas2014static}. However, each study only used a small number of features that are either easily collected or publicly published by existing work. 
Also, most existing works evaluated the proposed TCP techniques based on subjects with very low numbers of failed test cases (e.g., 2\% failure rate across subjects in \cite{bertolinolearning}) and short regression testing time (average regression testing below 90 seconds). Having a sufficient number of failed builds is important for evaluating such TCP techniques and creating the balanced datasets required by many ML techniques. Further, applying TCP techniques to systems whose regression testing takes a short time is not practically beneficial and therefore not representative in an experimental context. Our study focuses on these three mentioned issues by investigating the use of a comprehensive feature set for training ML TCP models based on 25 realistic subjects whose test failure rates have a median of 14\% and their regression testing time takes at least 5 minutes with a median of 12 minutes. In the following, we discuss the three studies closest to ours.

Bertolino et al.~\cite{bertolinolearning} analyzed the performance of ten ML algorithms, including three RL algorithms, for test prioritization in CI. Through experimental analysis, they showed that RL-based approaches to test case prioritization can adapt to new changes as their TCP effectiveness is less affected by changes. They also reported that MART~\cite{Ranklibs}, a pairwise ranking model that is trained using an ensemble of boosted regression trees, reaches the highest effectiveness among the ten ML and RL techniques.
However, their study suffers from the issues discussed above. More specifically, they only use limited features that are defined based on the execution history of test cases and their complexities. They also evaluated their work only based on six subjects, all of which include very low numbers of failed test cases (only 49 failed cycles among 2.6k builds across all subjects) and short regression testing time (maximum of 22 seconds across subjects). 

In a similar work to the previous one, Bagherzadeh et al.~\cite{ourRL} conducted a thorough analysis of ten state-of-the-art deep reinforcement learning techniques in the TCP context. Through extensive empirical analysis, they showed that the best result from RL techniques can reach similar effectiveness as MART in the TCP context. This study is based on the same data set as that of Bertolino et al.~\cite{bertolinolearning}. Thus it suffers from the same issues (limited number of features and subjects) as discussed above.

Elsner et al.~\cite{newbaseline}, investigated a more comprehensive set of features for training ML models in the TCP context. They found that features from test execution history work better than other types of features. They also showed that well-known and simple heuristics based on the failure rate of test cases often outperform complex ML models, which was unexpected.
While the paper used a relatively large number of subjects with an adequate number of failed test cases, they only used subjects containing only 16 features and thus this study is far from being comprehensive. Indeed, it missed many essential features, including features based on static coverage analysis and the code complexity of test cases and the SUT. Our study shows that complexity features play an important role in training highly effective ML models for TCP. They also did not investigate the data collection time associated with features, which is necessary to perform trade-off analysis on the cost and benefits of features. Finally, they failed to use the state-of-the-art ML models in a TCP  context (MART) as reported by previous studies~\cite{bertolinolearning}. Instead, they trained a point-wise ranking model that, according to existing studies (Bertolino et al.~\cite{bertolinolearning} and Bagherzadeh et al.~\cite{ourRL}), provides lower effectiveness compared to pairwise ranking models such as MART. We believe that all these elements explain why the effectiveness of their ML models is lower than simple heuristics. We further investigated (Section~\ref{sec:validation}) this issue by comparing the results of MART models against the heuristics based on failure rate, which reaches the best effectiveness as reported by Elsner et al.~\cite{newbaseline}. The results show that the models trained either based on the full feature set or features related to the execution history outperform the heuristic model.

Overall, our work complements existing ML-based TCP studies by performing a comprehensive investigation of features (150), obtained from various sources, for training ML TCP models. This allows us to provide practical insights on the benefits and costs of ML-based techniques through extensive empirical analysis. Also, we provide a tool and a set of diverse subjects that can be used as a benchmark for future studies.

\section{Conclusion}
\label{sec:conclusion}
In this work, focused on ML-based Test Case Prioritization (TCP), we have carefully defined a comprehensive set of features aimed at predicting the probability of regression failure, based on related studies, and a data model that captures entities and their relations in a typical CI environment. The features are categorized according to common data collection sources and steps into three high-level groups: \textit{REC} containing features capturing the test execution history, \textit{TES} containing features characterizing the complexity of the test case source code, their changes, and their development process, and \textit{COV} containing features that capture test case code coverage of changes in the system under test (SUT).
We have developed a tool to collect these features for 25 carefully selected open-source projects in order to carry out an extensive experimental study to analyze the cost and benefits of the feature groups for ML-based TCP. The ultimate goal is to provide concrete and precise recommendations regarding what data to collect to effectively support TCP in a scalable way.

The results of our study show that:
\begin{itemize}
    \item The data collection time of all features ranges between 0.1 to 11.7 minutes across subjects for each build. Also, the trained models for TCP based on the full feature set can achieve promising results in terms of $\text{APFD}_C$ across most subjects, with an average of 0.82.
    \item Coverage-related features (i.e., feature groups with \textit{COV} in their name) are the most expensive features to collect, while they have the least impact on the effectiveness of ML TCP models.
    \item \textit{REC} features are the least expensive to collect, while models relying on them achieve an $\text{APFD}_C$ close to models trained with the full set of features.
    \item Not retraining models for more than 11 builds leads to unstable $\text{APFD}_C$. Also, to achieve the best TCP effectiveness, the models should be retrained as frequently as possible.
\end{itemize}
 
Access to high-quality datasets is one of the main issues for devising and evaluating new TCP techniques. Thus, we also made our datasets publicly available, which can serve as a benchmark for future studies. In addition, we plan to extend our data collection tool to support the data collection at the method level and investigate how using method-level data can impact the TCP techniques in terms of data collection cost and effectiveness. Further, extending the data collection and analysis to other CI tools such as GitHub Actions and other programming languages would be a useful endeavor.

\section*{Acknowledgement}
This work was supported by a research grant from Huawei Technologies Canada, Mitacs Canada, as well as the Canada Research Chair and Discovery Grant programs of the Natural Sciences and Engineering Research Council of Canada (NSERC). This research was also enabled in part by the computation support provided by Compute Canada\footnote{\underline{\url{www.computecanada.ca}}}.

\bibliographystyle{IEEEtran}
\bibliography{main.bib}

\clearpage
\begin{appendices}
\section{Metric Definitions}
\label{appendix1}

\begin{table*}[ht!]
\centering
\caption{Test case feature groups.}
\label{tab:detailsfeatures}
\begin{tabular}{p{3cm}p{14.5cm}}
\hline
\small \textbf{Group(s)} & \small \textbf{Description}  \\ 
\hline
\textbf{TES\_COM}, \textbf{TES\_PRO}, \textbf{TES\_CHN} & Features characterizing the source code of test cases that are calculated based on code metrics (complexity, process, and change metrics).  \\ \hline 
\small \textbf{REC} & Features characterizing test cases based on their execution records (time and verdict) in builds. \\ \hline
\small \textbf{F\_COV} & Features capturing the file coverage of test cases in changed and impacted files. \\ \hline
\textbf{COD\_COV\_COM}, \textbf{COD\_COV\_PRO}, \textbf{COD\_COV\_CHN} & Features characterizing the source code of the changed and impacted files covered by test cases, which are calculated based on code metrics (complexity, process, and change metrics). \\ \hline
\small \textbf{DET\_COV} & Features based on the history of faults present in source files and detected by test cases. \\ \hline
\end{tabular}
\end{table*}

\begin{table*}[ht!]
\centering
\caption{Complexity, process, and change metrics of source code.} 
\label{tab:metrics}
\begin{tabular}{p{1.5cm}p{11.2cm}p{4.3cm}}
\hline
 \small \textbf{Group} &      \small \textbf{Metrics}  &      \small \textbf{Description}   \\ 
\hline
\small \textbf{Complexity (COM)} &\small \textbf{Program Size}: CntDeclFunction, LoC, LoCBlank, LoCCode,
LoCCodeDecl, LoCCodeExe, LoCComment, CntStmt, CntStmtDecl, CntStmtExe,
RatioCommentToCode \newline
\textbf{McCabe’s Cyclomatic Complexity}: MaxCyclomatic, MaxCyclomaticModified, MaxCyclomaticStrict, MaxEssential, MaxNesting, SumCyclomatic, SumCyclomaticModified, SumCyclomaticStrict, SumEssential \newline
\textbf{Object-oriented Metrics}: CntDeclClass, CntDeclClassMethod, CntDeclClassVariable, CntDeclExecutableUnit, CntDeclInstanceMethod,
CntDeclInstanceVariable, CntDeclMethod, CntDeclMethodDefault, CntDeclMethodPrivate, CntDeclMethodProtected, CntDeclMethodPublic & \small This metric group includes: \newline
1) Size metrics related to the number of lines of code, declarations, statements, and files. \newline
2) Complexity metrics related to the control flow graph of methods. \newline
3) Metrics based on object-oriented constructs. \\ \hline     
\small \textbf{Process (PRO)} & \small CommitCnt, DistinctDevCnt, OwnerContributedLines, MinorContributorCnt, OwnerExperience, AllCommittersExperience & \small Metrics based on the history of the development process of files in the SUT.  \\ \hline
\small \textbf{Change (CHN)} & \small LinesAdded, LinesDeleted, ChangedCodeScattering, DMMUnitComplexity, DMMUnitInterfacing, DMMUnitSize & \small Metrics focused on changes of the last version of the system under test. \\ \hline
\end{tabular}
\end{table*}

In this section, we provide definitions of the metrics used in our test case feature model. We use the standard definitions from the relevant sources \cite{understand, ProcessMetrics,  PyDriller, DMM, CodeScattering} and highlight and justify our definitions when they deviate from the standard ones. The metrics are defined based on a source code file. However, similar definitions can be provided based on a class or a method. In Table~\ref{tab:detailsfeatures} and Table~\ref{tab:metrics}, we also provide summarized descriptions of the feature groups and the metrics, respectively. 

\subsection{Complexity Metrics}
\subsubsection{Program Size}
\begin{itemize}
    \item \textbf{CntDeclFunction:} Number of functions in a file.
	\item \textbf{LoC:} Number of all lines in a file. [aka NL]
    \item \textbf{LoCBlank:} Number of blank lines in a file. [aka BLoC]
    \item \textbf{LoCCode:} Number of lines containing source code in a file. [aka LoC]
    \item \textbf{LoCCodeDecl:} Number of lines containing declarative source code in a file.
    \item \textbf{LoCCodeExe:} Number of lines containing executable source code in a file.
    \item \textbf{LoCComment:} Number of lines containing comments in a file. [aka CLoC]
    \item \textbf{CntStmt:} Number of statements in a file.
    \item \textbf{CntStmtDecl:} Number of declarative statements in a file.
    \item \textbf{CntStmtExe:} Number of executable statements in a file.
    \item \textbf{RatioCommentToCode:} Ratio of comment lines to code lines in a file.
\end{itemize}

\subsubsection{McCabe’s Cyclomatic Complexity}
\begin{itemize}
	\item \textbf{MaxCyclomatic:} Maximum cyclomatic complexity\footnote{\url{https://support.scitools.com/t/understanding-mccabe-cyclomatic-complexity/69}} of all nested functions or methods in a file.
    \item \textbf{MaxCyclomaticModified:} Maximum modified cyclomatic complexity of nested functions or methods in a file.
    \item \textbf{MaxCyclomaticStrict:} Maximum strict cyclomatic complexity of nested functions or methods in a file.
    \item \textbf{MaxEssential:} Maximum essential complexity of all nested functions or methods in a file.
    \item \textbf{MaxNesting:} Maximum nesting level of control constructs in a file.
    \item \textbf{SumCyclomatic:} Sum of cyclomatic complexity of all nested functions or methods in a file. [aka WMC]
    \item \textbf{SumCyclomaticModified:} Sum of modified cyclomatic complexity of all nested functions or methods in a file.
    \item \textbf{SumCyclomaticStrict:} Sum of strict cyclomatic complexity of all nested functions or methods in a file.
    \item \textbf{SumEssential:} Sum of essential complexity of all nested functions or methods in a file.
\end{itemize}

\subsubsection{Object-oriented Metrics}
\begin{itemize}
	\item \textbf{CntDeclClass:} Number of classes in a file.
	\item \textbf{CntDeclClassMethod:} Number of class methods in a file.
	\item \textbf{CntDeclClassVariable:} Number of class variables in a file.
	\item \textbf{CntDeclExecutableUnit:} Number of program units with executable code in a file.
	\item \textbf{CntDeclInstanceMethod:} Number of instance methods in a file. [aka NIM]
	\item \textbf{CntDeclInstanceVariable:} Number of instance variables in a file. [aka NIV]
	\item \textbf{CntDeclMethod:} Number of local methods in a file.
	\item \textbf{CntDeclMethodDefault:} Number of local default methods in a file.
    \item \textbf{CntDeclMethodPrivate:} Number of local private methods in a file. [aka NPM]
    \item \textbf{CntDeclMethodProtected:} Number of local protected methods in a file.
    \item \textbf{CntDeclMethodPublic:} Number of local public methods in a file. [aka NPRM]
\end{itemize}

\subsection{Process Metrics}
\begin{itemize}
	\item \textbf{Commit Count:} The number of commits made to a file.
	\item \textbf{Distinct Dev Cnt:} The cumulative number of distinct developers who contributed to this file up to this build/release.
	\item \textbf{Owner’s Contributed Lines:} The percentage of lines authored by the highest contributor of a file (the contributor with the most authored lines). Both added and deleted lines are counted as authored lines.
	\item \textbf{Minor Contributor Cnt:} The number of contributors who authored less than 5\% of the code in the file.
	\item \textbf{Owner’s Experience:} Measures the experience of the highest contributor of the file using the percent of lines he/she authored in the project up to this build/release.
	\item \textbf{All Committer’s Experience:} The geometric mean of experience of all the developers of the file.
	
\end{itemize}

\subsection{Change Metrics}
\begin{itemize}
\item \textbf{Lines Added and Deleted:} The added and deleted lines in the file in this build.
\item \textbf{Added and Deleted Change Scattering:} Given the added and deleted changes of a source file in a build, this metric measures the scattering of changes in the file. Assuming \(CH = \{ch_1, ch_2, ..., ch_n\}\) is the set of added/deleted code chunks in file $f$, we define Change Scattering (CS) as follows.
	\[CS(f) = \frac{|CH|}{\binom{|CH|}{2}}\times\sum\limits_{\forall ch_i, ch_j \in CH}\text{dist}(ch_i, ch_j)\]
	where $\text{dist}(ch_i, ch_j)$ computes the distance between the two code chunks $ch_i$ and $ch_j$ and is defined as follows.
	\[\text{dist}(ch_i, ch_j) = |line(ch_i) - line(ch_j)|\]
	where $\text{line}(ch_i)$ computes line number of the beginning of code chunk $ch_i$. If there is only one code chunk in the changed file $f$ (i.e., $|CH| = 1$), then $CS(f) = 0$. 
	
	The multiplication factor at the beginning of CS's formula has two objectives: (i) normalizing the distances between the code chunks by the number of pairs of code chunks modified in a build (reflected in the denominator), (ii) assigning a higher scattering to files including a higher number of changed code chunks (reflected in the numerator).
	
	This metric is inspired by previous work~\cite{CodeScattering} which proposes structural scattering for computing a developer's change scattering in a given period of time.
	\item \textbf{Delta Maintainability Metric (DMM)}: In one sentence, DMM is the proportion of low-risk change in a commit. The resulting value ranges from 0.0 (all changes are risky) to 1.0 (all changes have a low risk). DMM was originally calculated at the method level, and to calculate it at the commit level, we use \[DMM = \frac{lr\_chn}{lr\_chn + hr\_chn}\] where lr\_chn and hr\_chn stand for the number of high-risk and low-risk changes, respectively. A low-risk change is defined as adding low-risk code or removing high-risk code, whereas a high-risk change is defined as adding high-risk code or removing low-risk code.
    
    The DMM can be used on arbitrary properties that can be determined at the method (unit) level. The PyDriller \cite{PyDriller} OS-DMM implementation supports three properties:
    \begin{itemize}
        \item \textbf{DMMUnitComplexity:}  Cyclomatic complexity of method; low risk threshold 5.
        \item \textbf{DMMUnitInterfacing:} The number of method's parameters: low risk threshold 2.
        \item \textbf{DMMUnitSize:} Method size in lines of code; low risk threshold 15.
    \end{itemize}
    For instance, removing lines of code from a 20-line-method would be a low-risk change. However, adding lines of code to a method with three parameters counts as a high-risk change.
\end{itemize}

\section{Usage Frequencies of All Features}

\TR{R2.11}{As a supplement to research question 2.4, which is discussed and analyzed in Section 4.4.2 of the paper, Table~\ref{tab:rq2-fstats-all} shows the average usage frequency for all individual features in this study across all trained RF ranking models for all builds.}

\begin{table*}[ht]
\caption{\TRC{Usage frequency of individual features in models that were trained on the full feature set. The \textit{Avg. Freq.} column shows the average usage frequency of a feature across all trained RF ranking models for all builds.}}
\label{tab:rq2-fstats-all}
\centering
\resizebox*{!}{\textheight}{%
\begin{tabular}{llr|llr}
\toprule
Feature Group &                Feature &  Avg. Freq. & Feature Group &                Feature &  Avg. Freq. \\
\midrule
           REC &                        Age &       17034 &  COD\_COV\_COM &             CountDeclClass &         309 \\
      TES\_PRO &           OwnersExperience &       10618 &  COD\_COV\_COM &          CountLineCodeDecl &         302 \\
      TES\_PRO &     AllCommitersExperience &        7643 &  COD\_COV\_COM &             CountLineBlank &         293 \\
           REC &            TotalMaxExeTime &        5409 &  COD\_COV\_COM &      CountDeclMethodPublic &         292 \\
           REC &                LastExeTime &        4095 &  COD\_COV\_COM &              CountStmtDecl &         291 \\
      TES\_PRO &         OwnersContribution &        3997 &  COD\_COV\_COM &          CountDeclFunction &         287 \\
      TES\_PRO &                CommitCount &        3986 &  COD\_COV\_COM &                  CountLine &         277 \\
           REC &            TotalAvgExeTime &        3977 &  COD\_COV\_COM &            CountDeclMethod &         273 \\
           REC &           RecentAvgExeTime &        3850 &  COD\_COV\_COM &           CountLineCodeExe &         272 \\
           REC &           RecentMaxExeTime &        3741 &            REC &              RecentExcRate &         270 \\
      TES\_COM &         RatioCommentToCode &        3695 &  COD\_COV\_COM &    CountDeclExecutableUnit &         261 \\
      TES\_COM &              CountStmtDecl &        3383 &            COV &                ChnScoreSum &         261 \\
      TES\_COM &          CountLineCodeDecl &        3342 &  COD\_COV\_COM &               SumEssential &         255 \\
      TES\_COM &             CountLineBlank &        3149 &  COD\_COV\_COM &              CountLineCode &         248 \\
      TES\_COM &               CountStmtExe &        2825 &  COD\_COV\_COM &               CountStmtExe &         248 \\
      TES\_COM &                  CountLine &        2792 &  COD\_COV\_COM &                  CountStmt &         248 \\
      TES\_COM &           CountLineCodeExe &        2688 &  COD\_COV\_PRO &     AllCommitersExperience &         243 \\
      TES\_COM &           CountLineComment &        2668 &  COD\_COV\_COM &        SumCyclomaticStrict &         238 \\
      TES\_COM &                  CountStmt &        2658 &       TES\_COM &     CountDeclMethodDefault &         238 \\
      TES\_COM &              CountLineCode &        2453 &  COD\_COV\_COM &              SumCyclomatic &         230 \\
      TES\_COM &      CountDeclMethodPublic &        1529 &            REC &           RecentAssertRate &         230 \\
      TES\_COM &  CountDeclInstanceVariable &        1489 &  COD\_COV\_COM &      SumCyclomaticModified &         228 \\
      TES\_PRO &           DistinctDevCount &        1374 &  COD\_COV\_PRO &           OwnersExperience &         228 \\
      TES\_COM &        SumCyclomaticStrict &        1315 &  COD\_COV\_CHN &                 LinesAdded &         224 \\
           REC &          LastTransitionAge &        1284 &  COD\_COV\_COM &         RatioCommentToCode &         207 \\
      TES\_COM &              SumCyclomatic &        1281 &  COD\_COV\_PRO &         OwnersContribution &         196 \\
      TES\_COM &      SumCyclomaticModified &        1280 &  COD\_COV\_CHN &               LinesDeleted &         192 \\
      TES\_COM &    CountDeclInstanceMethod &        1273 &  COD\_COV\_CHN &      AddedChangeScattering &         168 \\
           REC &             LastFailureAge &        1268 &       TES\_COM &               MaxEssential &         157 \\
           REC &        TotalTransitionRate &        1141 &  COD\_COV\_CHN &    DeletedChangeScattering &         150 \\
      TES\_PRO &      MinorContributorCount &        1134 &  COD\_COV\_COM &     CountDeclClassVariable &         131 \\
      TES\_COM &            CountDeclMethod &        1068 &  COD\_COV\_COM &                 MaxNesting &         130 \\
           REC &              TotalFailRate &        1058 &  COD\_COV\_COM &  CountDeclInstanceVariable &         128 \\
      TES\_COM &               SumEssential &        1055 &       DET\_COV &                     Faults &         126 \\
      TES\_COM &    CountDeclExecutableUnit &        1039 &  COD\_COV\_PRO &                CommitCount &         125 \\
      TES\_COM &          CountDeclFunction &        1035 &  COD\_COV\_COM &               MaxEssential &         124 \\
      TES\_COM &     CountDeclClassVariable &         995 &  COD\_COV\_CHN &                    DMMSize &         119 \\
 COD\_COV\_PRO &           OwnersExperience &         971 &  COD\_COV\_COM &           CountLineComment &         119 \\
 COD\_COV\_PRO &     AllCommitersExperience &         908 &  COD\_COV\_COM &       CountDeclClassMethod &         119 \\
      TES\_COM &        MaxCyclomaticStrict &         845 &  COD\_COV\_COM &        MaxCyclomaticStrict &         115 \\
           REC &               TotalExcRate &         843 &            COV &                   ChnCount &         115 \\
 COD\_COV\_COM &         RatioCommentToCode &         797 &  COD\_COV\_COM &      MaxCyclomaticModified &         114 \\
      TES\_COM &              MaxCyclomatic &         784 &  COD\_COV\_COM &              MaxCyclomatic &         114 \\
      TES\_COM &     CountDeclMethodPrivate &         777 &  COD\_COV\_COM &     CountDeclMethodPrivate &         112 \\
      TES\_COM &      MaxCyclomaticModified &         769 &  COD\_COV\_COM &    CountDeclInstanceMethod &         108 \\
           COV &                ImpScoreSum &         761 &       TES\_CHN &                 LinesAdded &         108 \\
      TES\_COM &             CountDeclClass &         749 &  COD\_COV\_CHN &              DMMComplexity &         107 \\
      TES\_COM &                 MaxNesting &         740 &  COD\_COV\_PRO &           DistinctDevCount &         105 \\
      TES\_COM &       CountDeclClassMethod &         661 &  COD\_COV\_PRO &      MinorContributorCount &         105 \\
           REC &        MaxTestFileFailRate &         645 &  COD\_COV\_COM &             CountDeclClass &         101 \\
           REC &            TotalAssertRate &         612 &  COD\_COV\_COM &      CountDeclMethodPublic &         100 \\
           REC &  MaxTestFileTransitionRate &         597 &       TES\_CHN &               LinesDeleted &         100 \\
 COD\_COV\_PRO &         OwnersContribution &         597 &  COD\_COV\_COM &     CountDeclMethodDefault &          96 \\
           COV &                   ImpCount &         507 &  COD\_COV\_CHN &             DMMInterfacing &          94 \\
 COD\_COV\_PRO &                CommitCount &         458 &  COD\_COV\_COM &   CountDeclMethodProtected &          94 \\
 COD\_COV\_COM &     CountDeclClassVariable &         455 &  COD\_COV\_COM &           CountLineCodeExe &          93 \\
 COD\_COV\_COM &     CountDeclMethodDefault &         455 &  COD\_COV\_COM &          CountLineCodeDecl &          92 \\
 COD\_COV\_COM &       CountDeclClassMethod &         451 &  COD\_COV\_COM &                  CountLine &          90 \\
           REC &                LastVerdict &         443 &  COD\_COV\_COM &               SumEssential &          90 \\
           REC &             RecentFailRate &         441 &  COD\_COV\_COM &               CountStmtExe &          88 \\
 COD\_COV\_COM &  CountDeclInstanceVariable &         437 &  COD\_COV\_COM &             CountLineBlank &          88 \\
 COD\_COV\_COM &           CountLineComment &         418 &  COD\_COV\_COM &              CountLineCode &          87 \\
      DET\_COV &                     Faults &         403 &  COD\_COV\_COM &        SumCyclomaticStrict &          87 \\
 COD\_COV\_PRO &           DistinctDevCount &         389 &  COD\_COV\_COM &          CountDeclFunction &          87 \\
 COD\_COV\_PRO &      MinorContributorCount &         386 &  COD\_COV\_COM &    CountDeclExecutableUnit &          86 \\
      TES\_COM &   CountDeclMethodProtected &         360 &  COD\_COV\_COM &              CountStmtDecl &          86 \\
 COD\_COV\_COM &    CountDeclInstanceMethod &         359 &  COD\_COV\_COM &                  CountStmt &          86 \\
           REC &       RecentTransitionRate &         355 &  COD\_COV\_COM &              SumCyclomatic &          85 \\
 COD\_COV\_COM &               MaxEssential &         352 &  COD\_COV\_COM &            CountDeclMethod &          85 \\
 COD\_COV\_COM &     CountDeclMethodPrivate &         351 &       TES\_CHN &      AddedChangeScattering &          85 \\
 COD\_COV\_COM &        MaxCyclomaticStrict &         332 &  COD\_COV\_COM &      SumCyclomaticModified &          84 \\
 COD\_COV\_COM &                 MaxNesting &         331 &       TES\_CHN &    DeletedChangeScattering &          78 \\
 COD\_COV\_COM &              MaxCyclomatic &         325 &       TES\_CHN &                    DMMSize &          70 \\
 COD\_COV\_COM &   CountDeclMethodProtected &         317 &       TES\_CHN &             DMMInterfacing &          47 \\
 COD\_COV\_COM &      MaxCyclomaticModified &         309 &       TES\_CHN &              DMMComplexity &          46 \\
\bottomrule
\end{tabular}
}
\end{table*}

\end{appendices}

\end{document}